\documentclass[11pt]{article}
\usepackage{epsfig,enumerate,verbatim}
\usepackage{amsmath}
\usepackage{amsfonts}
\usepackage{amssymb}
\usepackage{amstext}
\usepackage{amsthm}
\usepackage{undertilde} % command is \utilde
\usepackage{dsfont}
\usepackage{graphicx}
\usepackage{mathrsfs}
\begin{document}        
\title{\bf{The Constraints and Spectra of a Deformed Quantum Mechanics}}

\author{Chee-Leong Ching\footnote{Email: phyccl@nus.edu.sg}, Rajesh R. Parwani\footnote{Email: parwani@nus.edu.sg} and Kuldip Singh\footnote{Email: cqtks@nus.edu.sg}}

\date{March 2012}

\maketitle

\begin{center}
{Department of Physics,\\}
{National University of Singapore,\\}
{Kent Ridge,\\}
{Singapore.}
\end{center}

\vspace{1.5cm}

\begin{abstract}
\noindent We examine a deformed quantum mechanics in which the commutator between coordinates and momenta is a function of momenta. The Jacobi identity constraint on a two-parameter class of such modified commutation relations (MCR's) shows that they encode an intrinsic maximum momentum; a sub-class of which also imply a minimum position uncertainty. Maximum momentum causes the bound state spectrum  of the one-dimensional harmonic oscillator to terminate at finite energy, whereby classical characteristics are observed for the studied cases. We then use a semi-classical analysis to discuss general concave potentials in one dimension and isotropic power-law potentials in higher dimensions. Among other conclusions, we find  that in a subset of the studied MCR's, the leading order energy shifts of bound states are of opposite sign compared to those obtained using string-theory motivated MCR's, and thus these two cases are more easily distinguishable in potential experiments.  
\end{abstract}


\section{Introduction}

Over the years, various approaches to unifying quantum theory with gravity have suggested the existence of a  minimum measurable length \cite{min,sabine}. For example, string theory suggests a generalised uncertainty principle (GUP) which implies a minimum uncertainty in position \cite{string}. The string GUP may be derived from a modification of the usual Heisenberg algebra: In one-dimensional space the suggestion of Kempf {\it et.al.} \cite{kempf} is
\begin{eqnarray}
[X, P]&=& i\hbar\left(1+\beta P^2\right)\label{string1} \, 
\end{eqnarray} 
where $\beta >0$ is the deformation parameter; the $D$-dimensional version is listed below in Eq.(\ref{stringD}). 

The implications of the modified algebra (\ref{string1}, \ref{stringD}) have been investigated in a large number of papers, particularly the deformation of  spectra of quantum mechanical systems and how such effects may potentially be detected in future experiments, see \cite{kempf,laynam,das1, rel-string, Pedram3,optics}, and the recent review \cite{sabine} for more references.
Though inspired originally by Planck scale physics, brane-world scenarios \cite{brane} suggest that MCR's such as (\ref{string1}), and others below, might be relevant at more accessible energies (see also \cite{das1}). 
 
Motivated by ideas of maximum momentum in deformed special relativity \cite{dsr}, in Ref.\cite{das2} Ali, Das and Vagenas (ADV) proposed an extension of (\ref{string1}) allowing for  linear momenta terms on the right-hand-side, generally  
\begin{eqnarray}
[X_{i},P_{j}]&=& i\hbar\Bigl(\delta_{ij}+\delta_{ij}\alpha_{1}P+\alpha_{2}\frac{P_{i}P_{j}}{P}+\beta_{1}\delta_{ij}P^{2}+\beta_{2}P_{i}P_{j}\Bigr),\label{1}
\end{eqnarray}
where $P^{2}=\sum_{i=1}^{3} P_{i}\ P^{i}$. Latin indices $(i,j)$ represent the spatial part of spacetime indices and the $(\alpha_s,\beta_s)$ with $s=1,2$ are deformation parameters, with $\beta \sim \alpha^2$ dimensionally.  

Some consequences of the relation (\ref{1}) have  been studied in the literature \cite{das2, das4,Pedram1,Pedram2}, but for particular choices of the parameters $\alpha_s, \beta_s$. In particular, the Jacobi identity constraint on (\ref{1}) was solved to leading order in $\alpha$ while the $\beta$ values were chosen freely \cite{das2}. 

In the first part of this paper we will investigate the impact of maximum momentum on the spectra of quantum mechanical systems\footnote{See also the recent papers \cite{Pedram4,mig}, the first of which also realises minimum position uncertainty.}. To this end, we begin in the next Section by implementing the Jacobi identity constraint exactly on a larger class of modified commutation relations (MCR's) than (\ref{1}).

We find a two-parameter class of exact MCR's which encode an intrinsic maximum momentum; a sub-class of which also imply a minimum position uncertainty.  One member of the exactly realised MCR's will turn out to have the form (\ref{1}) but with all parameters fixed in terms of a single deformation parameter.

The utility of the exactly realised MCR's is that they allow us to investigate the phenomenon of maximum momentum in a mathematically controlled manner: Since dimensionally $P_{max} \sim O(1/\alpha)$, this invalidates perturbative treatments in general. 
In Sect.(4) we solve exactly the one-dimensional harmonic oscillator problem for two cases which show a termination of the bound state spectrum at finite energy, followed by a continuum, in stark contrast to the usual undeformed case.  The position and momentum uncertainties of the upper-most bound states are seen to display classical characteristics. 

In the second part of this paper we switch our focus, determining the leading order energy shifts of bound states due to the obtained class of MCR's. Since quantum mechanics has been well tested, the relevant regime for such a task is $\alpha P \ll 1$. Thus starting from the exact MCR's, we deduce a two-parameter family of MCR's which have the same form as (\ref{1}) and which satisfy the Jacobi identity approximately, with errors quantified. The ADV choice of parameters \cite{das2} will be seen to be an example from this family. 

We determine the energy shifts in Sect.(5) using a semi-classical analysis which we justify in Sect.(6). The advantage of the semi-classical analysis is that it gives, with relatively little effort, the high energy spectrum for a larger class of potentials in diverse dimensions. This allows us to identify certain universal features of deformed spectra due to the obtained MCR's and to contrast the results with those due to the string theory inspired MCR's (\ref{string1}), and other non-polynomial versions. 

Relativistic corrections to energy shifts are also discussed in Sect.(5), while isotropic power-law potentials in higher dimensions are studied in Sect.(6). Various consistency checks are performed on the semi-classical results in Sects. (5,6), such as comparing them with known exact computations and perturbative expressions. 

We summarise our main results in the concluding section while the Appendices contain other details.

\section{The Jacobi Identity Constraint and Maximum Momentum}
Consider the general   
class of symmetric commutators
\begin{equation}
[X_i, P_j] = i \hbar \left[ F(P) \delta_{ij} + G(P) P_i P_j \right] \ . \label{class}
\end{equation}
Let derivatives be denoted as $F^{'} = dF/dP^2 = (dF/dP)/(2P)$; note that, as in \cite{das2}, we permit $F,G$ to depend on $P$ and not just $P^2$.

We assume a minimal extension of the Heisenberg algebra, leaving the following commutators unchanged 
\begin{equation}
[X_{i},X_{j}] = [P_{i},P_{j}]=0 \hspace{0.5cm} \mbox{for all $(i,j)$} \, . \label{commute}
\end{equation}
Consistency of (\ref{class},\ref{commute}) requires that the Jacobi identity be satisfied,
\begin{eqnarray}
[[X_{i},X_{j}],P_{k}]+[[X_{j},P_{k}],X_{i}]+[[P_{k},X_{i}],X_{j}]=0 \, , \label{jacobi}
\end{eqnarray}
as a result of which we find the constraint
\begin{equation}
G(P) = \frac{2 F F^{'} } { F-2F^{'} P^2} \, .
\end{equation}

Suppose now we demand the diagonal part of (\ref{class}) be at most quadratic,
\begin{equation}
F(P) = F_{2}(P) = 1 + \alpha_1 P + \beta_1 P^2 \, , \label{F2}
\end{equation}
then
\begin{equation}
G(P) =  G_{*}(P) \equiv \frac{2 F_2 F^{'}_{2} } { 1- \beta_1 P^2} \, . \label{G1} 
\end{equation}
For $\beta_1 >0 $, $G_{*}(P)$ has a pole where the commutator ({\ref{class}) diverges, implying an intrinsic maximum momentum. The example studied in \cite{Pedram4} exhibits such a scenario.

For $\beta_1 <0$, $F_2(P)$ and hence the commutator (\ref{class}) vanish at some momentum, implying a singularity in the weight factor and integrals such as (\ref{16}). Thus again we have an intrinsic maximum momentum but with different characteristics than the previous case: Now the limit appears to indicate an approach  to a classical phase \cite{dsr, mig, jiz}. 

For $\beta_1 \equiv 0$ the denominator in (\ref{G1}) trivialises and the final result for (\ref{class}) is 
\begin{eqnarray}
[X_{i},P_{j}]&=& i\hbar (1-\alpha P) \left[\delta_{ij}-\alpha {P_{i}P_{j} \over P} \right] \label{exact}  ,
\end{eqnarray}
where we have set $-\alpha_1 =\alpha >0$. This solution also has an intrinsic maximum momentum\footnote{The cut-off does exist even if $\alpha <0$ as one can see in dynamics restricted to a one-dimensional subspace where a particle explores the open line in momentum space.}, $P < 1/\alpha$. In Appendix A we show that an exact implementation of the Jacobi identity on the ansatz (\ref{1},\ref{commute})  gives the same solution (\ref{exact}). 

The implications of an intrinsic momentum cut-off\footnote{The interesting possibility of a continuation beyond this point will be discussed elsewhere.} will be seen in the following sections where we will also briefly review some other MCR's from the literature which involve a momentum cut-off \cite{dsr, Pedram4, mig, jiz, cutoff}. 

In passing, we mention another solution to (\ref{F2},\ref{G1}) obtained  by taking $\alpha_1=0$. It results in a MCR which is like the string motivated version (\ref{stringD}) but now within commuting space.

\subsection{Truncations}
For $\beta_1 \neq 0$, (\ref{G1}) is not a quadratic polynomial. If however one assumes  $|\beta_1| P^2 \ll 1$, then with re-labellings $\alpha_1 = -\alpha$, $\beta_1 = r \alpha^2$, $r$ dimensionless,  one obtains for (\ref{class})  
\begin{eqnarray}
[X_{i},P_{j}]_{r} & \approx & i\hbar \left[\delta_{ij}-\alpha \left(\delta_{ij} P + \frac{P_{i}P_{j}}{P}\right)+\alpha^{2}(\delta_{ij} r P^2+ (2r+1) P_{i}P_{j}) \right] \  \nonumber \\
&& \label{approxquad}
\end{eqnarray}
with terms of order $(r\alpha^3 p^3)$ ignored. Included in (\ref{approxquad}) is the possibility $\alpha \to 0, \ r\alpha^2 \to \beta \neq 0$ which corresponds to the leading $\beta^{'} = 2\beta$ version of the stringy MCR (\ref{stringD}).

In other words, for $r \neq 0$ the MCR's (\ref{approxquad}) form a two-parameter, $(\alpha,r)$, family which have at most quadratic momenta terms on the right and which satisfy the Jacobi identity {\it approximately}: the regime of validity is $|r| (\alpha P)^2 \ll 1$. Including $O(p^3)$ terms in (\ref{F2}) contributes only to the ignored terms in (\ref{approxquad}). 

As discussed in Appendix A, the ADV \cite{das2}   choice of parameters in the ansatz (\ref{1}) corresponds to the case $r=1$ in the approximate class of MCR's (\ref{approxquad}). 

The special case $r=0$ ($\beta_1=0$)  corresponds to the one-parameter deformed commutator (\ref{exact}) which is at most quadratic in momenta and which satisfies the Jacobi identity exactly. The simplicity and larger range of applicability of this case singles it out for special attention.

\subsection{One dimensional Subspace}
There are many experimental situations where dynamics is restricted to an essentially one dimensional subspace, such as the Penning trap discussed in \cite{laynam}. The one-dimensional projection of the the exact and approximate MCR's (\ref{exact},\ref{approxquad}) can be written together as 
\begin{eqnarray}
[X,P]&=& i\hbar\left(1 -2\alpha P+ q \alpha^2 P^2\right)\label{10d}
\end{eqnarray}    
where we have introduced the dimensionless parameter $q=3r+1$. For $q=1$ this relation is exact as it follows from (\ref{exact}) while for $q \neq 1$ it inherits the external $ |r| (\alpha P)^2 \ll 1$ limitations. The relation (\ref{string1}) is included as the limit $\alpha \to 0, \ q\alpha^2 \to \beta$.

In the literature \cite{das2,das4,Pedram1,Pedram2}, investigations of (\ref{10d}) with $\alpha \neq 0$ have so far been performed using the approximate relation \eqref{adv} of ADV. In the one-dimensional quantum mechanics this means that $q=4$ has been implicitly chosen rather than other values, such as $q=1$ which is implied by a projection of the exact three-dimensional MCR \eqref{exact}.  

In the following sections we will study (\ref{10d}) for generic values of $q$. It is convenient to define the deformation factor
\begin{equation}
f(P) \equiv 1 -2\alpha P+ q \alpha^2 P^2
\end{equation}
and write the one-dimensional relation as 
\begin{equation}
[X,P] = i\hbar f(P)  \label{1d} \, .
\end{equation}

For real momentum $p$ the polynomial $f(p)$ has roots at $(1\pm \sqrt{1-q} )/ (\alpha q)$. Thus for $q>1$ the roots are away from the real line and (\ref{1d}) is well defined. However for $q=1$ there is a doubly degenerate real root and a momentum cutoff  $p< 1/(q \alpha)$ is required. For $q<1$ there are two real roots except at the point $q=0$ where there is only a single real root at $1/(2\alpha)$.  As we shall see, the position and nature of these roots determines the qualitatively different features of the deformed spectrum; this will be particularly transparent in the semi-classical analysis of Sect.(5) where $1/f$ enters as a weight function in the integrals and is solely responsible for the deformed spectrum.   
 
Remember that in addition to the intrinsic momentum cut-offs coming from the roots of $f(p)$, for $q\neq1$ there are the external momentum limits $ |r| (\alpha p)^2 \ll 1$ on the validity of the MCR's. For some cases these are compatible, for example for $q=0$ (that is $|r| =1/3$)  the intrinsic cut-off $\alpha p < 1/2$ means that $ |r| (\alpha p)^2 < 1/12 \ll 1$ is satisfied.  For $q>1$ there are no intrinsic cut-offs (in the approximate forms (\ref{10d})) and so for such cases any discussion of the large momentum regime is to be understood as a formal comparison with other cases, such as $q=1$, where there is an intrinsic momentum cut-off.

\subsection{GUP, Minimal Length and Maximum Momentum}

We saw that an intrinsic maximum momentum, $O(1/\alpha) \sim O(1/\sqrt{\beta})$, is encoded in the class of exactly realised MCR's defined by (\ref{class},\ref{commute},\ref{F2},\ref{G1}), a scenario favoured by deformed special relativity \cite{dsr}. Consider now for simplicity a one-dimensional projection of that class,
\begin{eqnarray}
[X, P ] & =& i \hbar (F_2 + G_{*} P^2) \nonumber \\
&=& i \hbar \frac{(1+ \alpha_1 P + \beta_1 P^2)^2}{1-\beta_1 P^2} \label{gup1} \, . 
\end{eqnarray}
Working in the momentum representation of Sect.(3), then below the maximum momentum $P_{max} =1/\sqrt{\beta}$ the right side of (\ref{gup1}) is larger than $(1+\alpha_1 P + \beta_1 P^2) $ for the sub-class whose parameters $(\alpha_1, \beta_1)$  make $F_2 > 1 $ (for example when both $\alpha_1, \beta_1$ are positive). 
Defining as usual $(\Delta X)^2 = \langle X^2 \rangle - \langle X \rangle^2$, it follows that for that sub-class
\begin{equation}
\Delta X \Delta P \ \ge \ {1 \over 2} \langle[X,P]\rangle \ \ge \ {\hbar \over 2} ( 1 +  \alpha_1 \langle P \rangle +  \beta_1 \langle P^2 \rangle ) \   . \label{gup}
\end{equation}
This can be written as the GUP
\begin{equation}
{2 \Delta X \over \hbar} \ge \frac{ (1 +\alpha_1 \langle P \rangle /2)^2  }{ \Delta P  }  + \frac{ (\beta_1 - \alpha_{1}^{2}/4) \langle P \rangle^2  }{ \Delta P } +  \beta_1 (\Delta P)  \ . \label{gup2}
\end{equation}

For $\beta_1 >  \alpha_{1}^2  / 4 >0 $ the right side of (\ref{gup2}) is positive, increasing when $\Delta P$ tends to zero or infinity.  
Thus there exists a sub-class of the exactly realised MCR's for which $\Delta X$ has a nonzero minimum; such a minimum position uncertainty is usually associated with some fundamental length scale \cite{min, sabine, string} and originally motivated the search for deformed Heisenberg algebras that would realise such scenarios \cite{kempf,cutoff}. Simply on dimensional grounds, $(\Delta X)_{min} \sim O(\hbar \alpha)$.

Note that for $q \neq 1$ the truncated versions of the MCR's displayed in Eq.(\ref{10d}) are valid only in the $(\alpha P)^2 \ll 1$ regime, as derived earlier, and so maximum  momentum in the form of poles in the MCR's is no longer visible, though for $q \le 1$ the form where the MCR's vanish at some intrinsic maximum momentum is realised. Indeed we will see in Sect.(4) an example with $\Delta X = \Delta P =0$. 

For the truncated versions (\ref{10d}),  $q>1$ seems to allow for a minimum position uncertainty but no intrinsic momentum cut-off while for $q \le 1$  we have a maximum momentum cut-off but apparently no minimum position uncertainty.

\section{The Momentum Representation}
The exact spectrum of the harmonic oscillator with the string theory inspired MCR (\ref{string1}) has been studied in detail in Refs.\cite{kempf, laynam, Pedram3}: In one dimension the position and momentum operators may be represented in momentum space by $P=p$, \ $X=(1+\beta p^2) x$  where $x,p$ are the canonical variables satisfying the usual Heisenberg algebra 
\begin{eqnarray}
[x,p]=i\hbar. \label{canon}
\end{eqnarray}
with $ x= i \hbar \frac{\partial}{\partial p}$.

In this Section and the next we will study the one-dimensional harmonic oscillator under the MCR (\ref{10d}) using analogous techniques. Higher dimensional extensions are discussed in Sect.(6). 

\subsection{Hermiticity of Operators} 
A representation of the algebra (\ref{1d}) is 
\begin{eqnarray}
P&=&p\label{13a}\\
X&=&i\hbar f(p) \frac{\partial}{\partial p} = i \hbar (1 -2\alpha p+ q \alpha^2 p^2) \frac{\partial}{\partial p} \label{13b} \, .
\end{eqnarray} 
As noted earlier, we require $q >1$ to have well-defined measures and operators acting on the open line $-\infty < p < \infty$. For $q \le 1$ the intrinsic  momentum cut-off $p < p_{max}$ must be implemented.

Hermiticity of $X$ requires the weight function $1/f(p)$ in the inner product:
\begin{equation}
\langle \phi|\psi\rangle = \int \frac{dp}{(1- 2\alpha p+ q \alpha^2 p^2)} \ \phi^{*}(p) \psi(p) \, .\label{16}
\end{equation}
Then 
\begin{eqnarray}
\left(\langle \phi|X|\psi\rangle\right)^{*} &=&\langle \psi|X|\phi\rangle+B(p) \nonumber
\end{eqnarray}
where the boundary term  
\begin{eqnarray}
B_{1}(p)&\equiv&-i\hbar\bigl(\psi^{*}(p)\phi(p)\bigr)\Bigr|_{-\infty}^{p_{max}} \nonumber
\end{eqnarray} 
is required to vanish. The operator $P=p$ is manifestly Hermitian. 

For the Hamiltonian of the harmonic oscillator,
\begin{eqnarray}
H(X,P)=\frac{P^2}{2m}+ \frac{m \omega^2 X^2}{2} \, , \label{sho} 
\end{eqnarray}
since we are using the representation  (\ref{13a},\ref{13b}), the kinetic part is trivially Hermitian. For the potential part, we have 
\begin{eqnarray}
\left(\langle\phi|X^2|\psi\rangle\right)^{*} &=&\langle\psi|X^2|\phi\rangle+\tilde{B}(p), \nonumber
\end{eqnarray}
\noindent where again the boundary term
\begin{eqnarray}
\tilde{B}(p)& \equiv & f(p) \Bigl[\phi(p)\frac{\partial}{\partial p}\psi^{*}(p)-\psi^{*}(p)\frac{\partial}{\partial p}\phi(p)\Bigr]\Bigr|_{-\infty}^{p_{max}} \nonumber
\end{eqnarray}
is required to vanish. 

Thus with appropriate boundary conditions on the momentum space wavefunctions, $X$ and $H$ are Hermitian. In particular, the energy spectrum will be real and bounded below as one can verify for the explicit solutions we find later.

\subsection{Perturbative Results}
For later comparison with our exact and semi-classical calculations, we record here standard perturbative calculations of the energy shift for the problem defined by (\ref{1d},\ref{sho}). Such a calculation has been done for $q=4$ in Ref.\cite{Pedram1} and it can easily be adjusted to give the result for general $q$: 
\begin{eqnarray}
\Delta E_{n}^{pert} &=& {m \alpha^2 \over 2} \Bigl[ {q (\hbar \omega)^2 \over 4} + (q-4) (E_{n(0})^2 \Bigr]  + O(\alpha^3) \ , \label{pert} \\
E_{n(0)} &\equiv & (n + {1 \over 2}) \hbar \omega \, .
\end{eqnarray}
Notice that there is no $O(\alpha)$ correction: This vanishes because the unperturbed states are parity eigenstates while the $O(\alpha)$ perturbing Hamiltonian is parity odd \cite{das2,Pedram1}. 

\section{The ``$\rho$-representation" and exact results}

The Schrodinger equation for the harmonic oscillator (\ref{sho}) in the momentum representation is  
\begin{eqnarray}
\left[\frac{p^2}{(m\hbar\omega)^2} -\Bigl( f(p) \frac{\partial}{\partial p}\Bigr)^2\right]\Psi(p)&=&\frac{2E}{m(\hbar\omega)^2}\Psi(p).\label{23}
\end{eqnarray}
It is useful to change variables from $p$ to a new variable $\rho$  in such a way that the equation takes the canonical Schrodinger form. The change is defined by
\begin{equation}
f(p) \frac{\partial}{\partial p}  \equiv  \frac{\partial}{\partial \rho} \label{rhotrans} \\
\end{equation} 
and results in the Schrodinger equation 
\begin{eqnarray}
\left[ -\frac{\partial^2}{\partial \rho^2} +  V(\rho)  \right]\Psi(\rho)&=&\frac{2E}{m(\hbar\omega)^2} \Psi(\rho) \  \label{rhorep} 
\end{eqnarray}
with positive potential 
\begin{equation}
 V(\rho) \equiv \frac{p^2(\rho)}{(m\hbar\omega)^2} \, .
\end{equation}
The function $p^2(\rho)$ is determined by solving (\ref{rhotrans}). Notice that the weight function is removed from the integration measure, $\int {dp \over f} \to \int d\rho$.

\subsection{q $>$ 1}

For $q>1$ we have 
\begin{eqnarray}
V(\rho) &=& \frac{ \left[1 + (q-1) \tan\theta(\rho) \right]^2}{ (q \alpha m\hbar\omega)^2} \nonumber ,\\
\theta(\rho) &\equiv & \alpha \sqrt{q-1} \ (\rho + C)  \nonumber ,\\
C &\equiv & {1 \over \alpha \sqrt{q-1}} \arctan{1 \over 1-q} \nonumber , 
\end{eqnarray} 
with $ \rho_{1} \leq \rho \leq \rho_{2} $. The boundaries $(\rho_1,\rho_2)$ are located where $\theta(\rho) = \pm \pi/2$. Since the potential is confining, approaching infinity at the boundaries, the spectrum will be entirely discrete and unbounded above 
\cite{messiah}. As the potential walls approximate an infinite well at high energies, large eigenvalues will take the form $E_n \sim n^2$,  a conclusion which we shall confirm using a semi-classical analysis in Sect.(5). One may also use the perturbative results (\ref{pert}) for moderate values of $n$, when the perturbation is small: The $n$ dependence of the energy shifts for $q<4$ is negative. This later trend will also be seen in more generality in the semi-classical  analysis.

Thus for $q>1$ though the spectrum is deformed from that of the usual $\alpha=0$ oscillator, particularly at large energies, it is still purely discrete and unbounded above. The situation will be quite different for $q \le 1$ as we shall soon see.

We will refer to (\ref{rhorep}) as the ``$\rho$-representation" of the Schrodinger equation; it is actually a Fourier transform of the alternative representation of the algebra in which one takes $X=x$ as a $c$-number \cite{prep}. Although such a representation has been used before to solve the Schrodinger equation with MCR's, see for example \cite{laynam,Pedram3,mig}, it is useful to note that much information about the deformed spectrum may be obtained simply by analysing the potential $V(\rho)$, and then appealing to standard results for the conventional Schrodinger equation \cite{messiah}. We will illustrate this further below.

\subsection{q=0}
For $q=0$ we have the intrinsic momentum cut-off $p< 1/(2 \alpha)$. Proceeding as in the previous section, the Schrodinger equation in the $\rho$-representation, after some scaling to make $\rho$ dimensionless, is now
\begin{eqnarray}
\left[ -\frac{\partial^2}{\partial \rho^2} +  V(\rho)  \right]\Psi(\rho)&=& \epsilon \Psi(\rho)\ .  \nonumber
\end{eqnarray}
where
\begin{equation}
 V(\rho) \equiv \frac{(1-e^{\rho})^2}{\delta^4} \; , \hspace{0.5cm} -\infty<\rho<\infty \ , \label{well1}
\end{equation}
and the parameters are
\begin{eqnarray}
\delta&=&2\sqrt{m\hbar\omega} \ \alpha \nonumber\\
\epsilon &=& \frac{2E}{\hbar\omega \delta^2} \ .
\end{eqnarray}

The potential has one minimum, $V(\rho=0)=0$, approaches positive infinity as $\rho \to \infty$ and approaches $1/\delta^4$ as $\rho \to -\infty$. Thus the bound state spectrum has a maximum energy limited by the depth of the well, $\epsilon_{max} \le 1/\delta^4$, which  translates to $E_{max} \le 1/(8m\alpha^2)$. The total number of bound states below an energy $E$ is approximately proportional to the integral \cite{messiah}
\begin{equation}
\int \sqrt{E - V(\rho)} \ d\rho \label{bddest}
\end{equation}
evaluated between the classical turning points; it essentially counts the number of half-deBroglie wavelengths that can fit at energy level $E$. Since the potential well in (\ref{well1}) approaches the $\rho \to -\infty$ asymptote exponentially, the integral is finite for $E$ at the top of the well and hence the number of bound states supported by the well is finite. 

Although the bound state spectrum terminates at finite energy, there are still continuum states of higher energies: The momentum cut-off $p< 1/(2 \alpha)$ does not imply that all energies are bounded! This conclusion is manifest in the $\rho$-representation of the Schrodinger equation and illustrates again its utility (See also Eqs. ({\ref{28}-\ref{ansatz}) below).

In summary, the spectrum for the $q=0$, $\alpha$-deformed oscillator is dramatically different from the usual $\alpha=0$ case: The present spectrum consists of a finite number of bound states followed by a continuum.  

We proceed now to determine the bound state energies and wavefunctions explicitly. As the procedure to solve equations such as (\ref{rhorep}) is standard 
\cite{bender} we will just outline the main steps. Firstly, an asymptotic analysis identifies the dominant behaviour of square-integrable solutions at the two ends, and then one uses an ansatz for the wavefunction which includes that information to simplify the differential equation. With a further change of variables 
\begin{eqnarray}
\xi&=&{2\over \delta^2} e^{\rho} \ ;\hspace{0.5cm}0<\xi<\infty \ , \label{28}
\end{eqnarray}
which changes the weighted measure to $\int {d \xi \over 2\alpha \xi}$, and defining
\begin{eqnarray}
k  &\equiv &  \sqrt{ {1\over \delta^4} - \epsilon} \ > 0 \ ,   \label{kk} \\
\tilde{n} & \equiv & {1 \over \delta^2} \Bigl[ 1-(1-\delta^4 \epsilon)^{1/2} \Bigr] -{1 \over 2} \ , \label{tiln}
\end{eqnarray}
we obtain 
\begin{eqnarray}
\Psi(\xi) &=& e^{-\xi/2} f(\xi) \ \Bigl( {\delta^2 \xi \over 2} \Bigr)^{k},\label{ansatz}
\end{eqnarray}
where $f(\xi)$ is a remaining function that satisfies the equation
\begin{eqnarray}
\xi\frac{\partial^2 f(\xi)}{\partial\xi^2}+\Bigl(2k+1-\xi\Bigr)\frac{\partial f(\xi)}{\partial\xi} + \tilde{n}f(\xi)&=&0.\label{30}
\end{eqnarray} 
The positivity requirement on $k$ comes from the square-integrability condition at $\xi=0$ ($\rho=-\infty$) and we see that it implies an upper bound on the bound state energies identical to what was deduced earlier from the depth of the potential well,
\begin{equation}
E_n \le E_{max} = {1 \over 8 m \alpha^2} \ . \label{emax}
\end{equation} 

Equation \eqref{30} has two singularities, a regular singular point at $\xi=0$ and an essential singularity at $\xi=\infty$. The function  $f(\xi)$  may be expanded in a Frobenius series about $\xi=0$,
\begin{eqnarray}
f(\xi)&=& \xi^s \sum_{j=0}^{\infty} a_{j}\xi^{j} \nonumber \,. 
\end{eqnarray} 
The indicial equation that follows is
\begin{equation}
s (s+ 2k) =0 \, . \nonumber
\end{equation}
The solution with $s=-2k$ is not square-integrable at $\xi=0$ and is discarded. The remaining $s=0$ case leads to an infinite series which must be truncated as otherwise its growth would be exponential and again lead to an unnormalisable solution (at $\xi=\infty$). Truncation leads to the quantisation condition 
\begin{equation}
\tilde{n} =n , \;\; n=0,1,2... \nonumber
\end{equation}
and thus from (\ref{kk},\ref{tiln}) we obtain 
\begin{eqnarray}
E_{n}&=& E_{n(0)}  \left[1- 2m \alpha^2 E_{n(0)} \right]  \, ,  \label{32} \\
n & \le & n_{max} = { 1 \over 4\alpha^2 m \hbar \omega} - {1 \over 2} \ ,
\end{eqnarray}
If $n_{max}$ is not an integer then the largest bound state realised is for an integer less than or equal to the value indicated.  In fact $E_n$ in (\ref{32}) is a monotonically increasing function of $n$, reaching a stationary point at an integer near $n_{max}$ 
where  $E=E_{max}$ (\ref{emax}) is attained, see Fig.(1). 

Amusingly, the exact bound state spectrum for $q=0$  has only an $\alpha^2$ correction, which matches the leading order perturbative result (\ref{pert}). Remarkably, as we will shall see later,   a leading order semi-classical analysis also reproduces the exact spectrum\footnote{It turns out that (\ref{well1}) is simply the Morse potential, while for the $q=1$ case the problem is similar to the radial Schrodinger equation for a Coulomb potential with centrifugal barrier. Thus it appears that the deformed harmonic oscillator with MCR (\ref{10d}) is related to some solvable potentials in ordinary quantum mechanics. This link can perhaps be explored using the methods of SUSY quantum  mechanics as in Refs.\cite{factor}.}

\subsubsection{Expectation Values and Uncertainties}
We proceed to obtain the wavefunctions. The reduced Schrodinger equation is
\begin{eqnarray}
\xi\frac{\partial^2 f(\xi)}{\partial\xi^2}+\Bigl(2k_n +1-\xi\Bigr)\frac{\partial f(\xi)}{\partial\xi}+nf(\xi)&=&0;\hspace{0.1cm}n\ni Z^{+} .\label{33}
\end{eqnarray}
with solutions given by  associated Laguerre polynomials $f(\xi)=L_{n}^{2k_n}(\xi)$.  The normalized wave functions are
\begin{eqnarray}
\Psi_{n}^{2k}(\xi)&=&\sqrt{\frac{ 4 \alpha k_n n!}{\Gamma(2k_n+n+1)}} \exp(-\xi/2)\xi^{k_n} L_{n}^{2k_n}(\xi) \, . \label{39}
\end{eqnarray}    
Note that due to the deformation, the index $k_n$ is in general no longer an integer even though $n$ is, 
\begin{eqnarray}
k_{n}&=&  {1 \over \delta^2} - (n+1/2)   \ , \label{34} 
\end{eqnarray} 
which makes the wavefunctions non-analytic at $\xi=0$, that is near $p=p_{max}$. 

The various expectation values in an eigenstate $n$ may now be evaluated exactly. We obtain  
\begin{eqnarray}
\langle X\rangle&=&0    \ , \nonumber\\
\langle X^2\rangle&=& \frac {E_{n(0)}} {m \omega^2} \left[1-4m\alpha^2  E_{n(0)}  \right] \ , \nonumber\\
\langle P\rangle&=& 2 m \alpha E_{n(0)} \ , \nonumber\\
\langle P^2\rangle&=& m E_{n(0)} . \label{expect}
\end{eqnarray}

In the limit
\begin{eqnarray}
n & \to& n_{\text{max}} \equiv \frac{1}{4m\hbar\omega\alpha^2}-\frac{1}{2}\ ,  \label{maxn}
\end{eqnarray} 
we have $<P> \to 1/(2\alpha)$, which is precisely the momentum cut-off the $q=0$ theory.  The mean potential and kinetic energies of a state $n$,  
\begin{eqnarray}
\langle V\rangle &=&  \frac {E_{n(0)}} {2} \left[1-4m\alpha^2  E_{n(0)} \right]   \ ,    \nonumber\\
\langle T\rangle &=&  \frac {E_{n(0)}} {2}  \ , \nonumber 
\end{eqnarray}
sum to the full energy but their difference shows a clear deviation from the $\alpha=0$ Virial Theorem \cite{prep}. 

The uncertainties are
\begin{eqnarray}
(\Delta X)^2&=&(n+ \frac{1}{2})\frac{\hbar}{m\omega} \left[\frac{n_{max} - n} {n_{max}+ 1/2}  \right]  \ , \nonumber\\
(\Delta P)^2&=&(n+\frac{1}{2} ) m\hbar\omega   \left[\frac{n_{max} - n} {n_{max}+ 1/2}  \right]     \  ,         \nonumber\\
\Rightarrow (\Delta X) (\Delta P)&=&(n+\frac{1}{2} ) \hbar  \left[\frac{n_{max} - n} {n_{max}+ 1/2}  \right]   \  .    \label{uncert1}
\end{eqnarray}

The position and momentum uncertainties, and their product, increase with $n$ up to $n \approx n_{max}/2$ and then decline, vanishing at $n=n_{max}$ if $n_{max}$ is an integer. This trend should be contrasted to the usual $\alpha=0$ harmonic oscillator when the same quantities increase linearly with $n$. (If $n_{max}$ is not an integer then (\ref{uncert1}) reaches the limit $(\delta n_{max}) \hbar$ where $\delta n_{max}$ is the deviation of $n_{max}$ from the largest integer smaller than or equal to $n_{max}$.) 

Thus the upper most bound states appear to have classical characteristics, as might have been anticipated also from the commutator (\ref{10d}) which vanishes at $p= p_{max}=1/(2\alpha)$. 

Let us look at how the wavefunctions behave as the quantum number increases. For small $\alpha$ and small $n$, $ \langle P \rangle \sim 0$ as in usual quantum mechanics since the wavefunction does not feel the effects of a maximum momentum, leading to $(\Delta P)^2 \sim \langle P^2 \rangle $. Fig.(2) shows the probability density for the $n=10$ state which is already slightly asymmetrical, with the highest peak closer to $p=p_{max}$ (which corresponds to the variable $\xi =0$). Notice the $n=10$ state has ten nodes as expected \cite{messiah}. 

As $n$ increases, there are more nodes and peaks but the dominant peak moves towards $\xi=0 \ (p=p_{max})$. Furthermore the ratio of heights of the dominant peak to the other peaks increases. Figs.(3,4) show the upper and lower end of the probability density for the upper-most state $n=49$; note the different scales in the two figures, and also the scale in Fig.(2). (Also note that the effects of the measure $\sim \int {d\xi \over \xi}$ will accentuate the difference between the dominant and other peaks). 

For large $n \sim n_{max} \sim 1/\alpha^2$ there is one dominant sharp peak near $p_{max}$, leading to $ \langle P \rangle ^2 \sim \langle P^2 \rangle \sim p_{max}^{2}$ and thus $\Delta P \approx 0$.

\subsubsection{Realising $\Delta P=0$ at $p_{max}$}

In usual quantum mechanics with a symmetrical potential in one dimension, $\langle P \rangle =0$ for eigenstates as a result of their parity. For large $n$, the momentum-space probability density will have two large peaks at $p \sim \pm \sqrt{2mE_n}$, giving a large value for $(\Delta P)^2 = \langle P \rangle^2$, with the distance between the peaks increasing with $n$. 

However in a MCR quantum theory with an intrinsic maximum momentum the situation is different. We saw above cases where, due to the parity non-invariance of the MCR, there is one  $p_{max} >0$ which causes large $n$ bound states to accumulate there leading to $\Delta P \approx 0$.    

If the intrinsic maximum momentum is realised symmetrically in a MCR quantum mechanics, then large $n$ bound states should have two dominant peaks at $\pm p_{max}$. By parity the eigenstates will have $\langle P \rangle =0$. However consider a superposition  $\phi_1 + \lambda \phi_2$ of two bound states of opposite parity and essentially same energy near the top end of the deformed discrete spectrum: They will have same size dominant peaks and so for the superposition $\langle P \rangle^2 \approx 4\lambda^2 p_{max}^2$ while $\langle P^2 \rangle \approx (1+ \lambda^2) p_{max}^{2}$ leading to $(\Delta P)^2 /  \langle P^2 \rangle \approx 0$ for suitable $\lambda$. (If the superposed states do not have exactly the same energy then there will be very slow dispersion.)  

%Thus we conjecture that in three dimensional MCR's of the form (\ref{class}) which have an intrinsic maximum momentum, regardless of whether the maximum is implied by a pole or zero on the right of (\ref{class}), there exists some superposition of bound states for which $\langle P^2 \rangle \sim p_{max}^2$ with the ratio  

As for the value of $\Delta X$ when $\Delta P=0$, a consistent possibility is zero when $[X,P]=0$, leading to a seemingly classical phase;  but if the commutator diverges because of a pole realised maximum momentum, then $\Delta X$ must diverge  too, leading to some new ``super"-quantum phase.

\subsection{q=1}

The $q=1$ case corresponds to the exactly realised MCR. We have the intrinsic momentum cut-off $p< 1/\alpha$. Proceeding as before, the Schrodinger equation  is 
\begin{eqnarray}
\left[ -\frac{\partial^2}{\partial \rho^2} +  V(\rho)  \right]\Psi(\rho)&=& \epsilon \Psi(\rho)\ .  \nonumber
\end{eqnarray}
with
\begin{equation}
 V(\rho) \equiv \frac{(\rho-1)^2}{\delta_{1}^{4} \rho^2}  \; , \hspace{0.5cm} 0 <\rho<\infty \ , \label{well2}
\end{equation}
and the parameters are
\begin{eqnarray}
\delta_1&=&\sqrt{m\hbar\omega} \  \alpha \nonumber\\
\epsilon &=& \frac{2E}{\hbar\omega \delta^{2}_{1}} \ .
\end{eqnarray}

The potential has one minimum, $V(\rho=1)=0$, approaches positive infinity as $\rho \to 0$ and approaches $1/\delta^{4}_{1}$ as $\rho \to \infty$. The bound state spectrum again terminates at finite energy, limited by the depth of the well, $\epsilon_{max} \le 1/\delta^{4}_{1}$, which  translates to $E_{max} \le 1/(2m\alpha^2)$. Since the potential flattens out slowly, the integral in (\ref{bddest}) now diverges and so, in contrast to the $q=0$ case,  the total number of bound states supported by the finite-depth well is infinite. 

The Schrodinger equation may be solved exactly as before leading to 
\begin{eqnarray}
E_n &=& \frac{ E_{n(0)} \sqrt{1 + {\delta_{1}^4 \over 2} }  + {m\alpha^2 \over 2} \left[ (E_{n(0)})^2  + {(\hbar \omega)^2 \over 4} \right]} {\left[  \sqrt{1 + {\delta_{1}^4 \over 2}} +   m \alpha^2 E_{n(0)} \right]^2 } \ .
\end{eqnarray}
The energies reach $E_{max}$ as $n \to \infty$ where they match the semi-classical result given in Appendix B. The leading $\alpha^2$ correction agrees with perturbative calculations (\ref{pert}) and differs from the semi-classical result (\ref{semihar}) by a constant. 

Defining
\begin{eqnarray}
\tilde{a} & \equiv & \frac{1+\sqrt{1+4/\delta_{1}^4}}{2} \, , \nonumber \\
k_n& \equiv& \frac{1}{\delta_{1}^4 (n+\tilde{a})}  \, , \nonumber \\
\eta &=& 2 k_n \rho \, ,
\end{eqnarray}
gives the wavefunctions 
\begin{eqnarray}
\Psi_{n}^{(2\tilde{a}-1)}(\eta)&=&\sqrt{\frac{k_{n}\ n!\ \alpha}{(n+\tilde{a})\ \Gamma[n+2\tilde{a}]}} e^{-\eta/2}\eta^{\tilde{a}}L_{n}^{(2\tilde{a}-1)}(\eta).\label{wavef3}
\end{eqnarray}

The exact expectation values in an eigenstate $n$, and their leading order terms are given by
\begin{eqnarray}
\langle X\rangle&=&0\nonumber\\
\langle X^2\rangle&=&\frac{\left[n+1/2+\frac{\delta_{1}^2}{2\sqrt{\widetilde{\delta_{1}}}}\right]\alpha^2\hbar^2}{\delta_{1}^8\left[n+1/2+\frac{\sqrt{\widetilde{\delta_{1}}}}{2\delta_{1}^2}\right]^3}\nonumber\\
&=&(n+1/2)\frac{\hbar}{m\omega}-\frac{1}{2}\bigl[6n(n+1)+1\bigr]\alpha^2\hbar^2+O(\alpha^4)\nonumber\\
\langle P\rangle&=&\frac{1}{\alpha}-\frac{1}{\left[n+1/2+\frac{\sqrt{\widetilde{\delta_{1}}}}{2\delta_{1}^2}\right]^2\alpha\delta_{1}^4}\nonumber\\
&=&2(n+1/2)m\hbar\omega\alpha+O(\alpha^3)\nonumber\\
\langle P^2\rangle&=&\frac{2n^3\sqrt{\widetilde{\delta_{1}}} +3n^2\left[\widetilde{\delta_{1}}+\sqrt{\widetilde{\delta_{1}}}\right] +n\left[3\widetilde{\delta_{1}}+3\sqrt{\widetilde{\delta_{1}}} +2\sqrt{\frac{\widetilde{\delta_{1}}}{\delta_{1}^4}}\right]+\left[\widetilde{\delta_{1}} +\sqrt{\widetilde{\delta_{1}}}+\sqrt{\frac{\widetilde{\delta_{1}}}{\delta_{1}^4}}-1\right]}{2\alpha^2\sqrt{\widetilde{\delta_{1}}}\left[n+1/2+\frac{\sqrt{\widetilde{\delta_{1}}}}{2\delta_{1}^2}\right]^3}\nonumber\\
&=&(n+1/2)m\hbar\omega-\left[2n^3+3n^2+\frac{9n}{8}+\frac{1}{16}\right]m^3\hbar^3\omega^3\alpha^4+O(\alpha^6).\label{expect1}
\end{eqnarray}
where we have denoted $\widetilde{\delta_{1}}:=4+\delta_{1}^4$. At the end of the bound state spectrum, $n \to \infty$, we see that $<P> \to 1/\alpha$, the intrinsic momentum cut-off of the theory. 

Using the above expressions one may verify that the kinetic and potential energies add to $E_n$ and that 
\begin{eqnarray}
\frac{\langle V\rangle}{\langle T\rangle} &=&1-\frac{m\hbar\omega\alpha^2}{2(n+1/2)}\bigl[6n(n+1)+1\bigr]+O(\alpha^4).
\end{eqnarray}  

The momentum and position uncertainties may also be computed exactly,
\begin{eqnarray}
\frac{(\Delta X)^2}{\hbar/(m\omega)}&=&\frac{n\widetilde{\delta_{1}} +\frac{\delta_{1}^2}{2}\Bigl[\sqrt{\widetilde{\delta_{1}}}+\delta_{1}^2\Bigr]+2}{\delta_{1}^6\widetilde{\delta_{1}}\left[n+1/2 +\frac{\sqrt{\widetilde{\delta_{1}}}}{2\delta_{1}^2} \right]^3}\nonumber\\
&=&(n+1/2)-\frac{1}{2}\bigl[6n(n+1)+1\bigr]m\omega\hbar\alpha^2+O(\alpha^4)\nonumber\\
\frac{(\Delta P)^2}{m\hbar\omega}&=&\frac{2(n+ 1/2)}{\delta_{1}^8\sqrt{\widetilde{\delta_{1}}}\left[n+1/2 +\frac{\sqrt{\widetilde{\delta_{1}}}}{2\delta_{1}^2}\right]^4}\nonumber\\
&=&(n+1/2)-\frac{1}{4}(n+1/2)^2 m\hbar\omega\alpha^2+O(\alpha^4)\nonumber\\
\Rightarrow \frac{(\Delta X) (\Delta P)}{\hbar}&=&\frac{\sqrt{2(n+1/2)\left[n\widetilde{\delta_{1}} +\frac{\delta_{1}^2}{2}\Bigl[\sqrt{\widetilde{\delta_{1}}}+\delta_{1}^2\Bigr]+2\right]}}{\delta_{1}^{7}\widetilde{\delta_{1}}^{3/4}\left[n+1/2 +\frac{\sqrt{\widetilde{\delta_{1}}}}{2\delta_{1}^2} \right]^{7/2}}\nonumber\\
&=&(n+1/2)-\frac{\bigl[14n(n+1)+3\bigr]}{4} m\hbar\omega\alpha^2+O(\alpha^4).\label{uncert}
\end{eqnarray}

The leading order expressions for the uncertainties, which have also been indicated above,  show that they are smaller than the $\alpha=0$ case. 
The exact expressions are plotted  Figs.(2-4). As for the $q=0$ case they increase up to some value 
$n \sim O(1/(m \hbar \omega \alpha^2))$ after which they decline, vanishing at the end of the bound state spectrum, $n =\infty$, as one may verify from the exact expressions.

\section{Semi-classical Analysis}

In usual quantum mechanics, the Sommerfeld-Wilson semi-classical quantisation rule, the leading part of a $\hbar$ expansion, accurately describes  the high-energy bound state spectrum with relatively little effort. Such a semi-classical approach has been also been used  for deformed quantum mechanics of the type (\ref{string1}) \cite{Pedram3, Fityo} and the results have been shown to be in very good agreement with exact solutions for various cases.   

In the following, we adopt the semi-classical approach to determine the  bound state energies of one-dimensional potentials under the deformation (\ref{1d}). 
(A detailed explanation and justification of the semi-classical approximation in the case of deformed quantum mechanics is in Sect.(6), see also \cite{Fityo}).

\subsection{Harmonic Oscillator}
From the representation in \eqref{13a} and \eqref{13b}, the semi-classical energy can be written in terms of the canonical classical coordinates $(x,p)$ as, 
\begin{eqnarray}
E^{sc}&=&\frac{p^2}{2m}+\frac{m\omega^2}{2}\Bigl[1- 2\alpha p + q\alpha^2 p^2\Bigr]^2 x^2 \label{scE}\\
\Rightarrow\ x&=&\pm\sqrt{\frac{2m E^{sc}-p^2}{m^2\omega^2\bigl[1- 2\alpha p + q \alpha^2 p^2\bigr]^2}}\nonumber\\
&=&\pm\frac{\sqrt{z^2-p^2}}{m\omega\bigl[(1- 2\alpha p + q \alpha^2 p^2\bigr]} \, , \label{40}
\end{eqnarray}
with $z \equiv \sqrt{2m E^{sc}} >0$. Notice from (\ref{scE}) that $E^{sc} >0$ by construction. 

The phase-space area in the Sommerfeld-Wilson quantization rule
\begin{eqnarray}
\oint x\ dp=\bigl(n+1/2\bigr) h; \hspace{0.35cm} n=0,1,2,3....\label{BS}
\end{eqnarray} 
is  then
\begin{eqnarray}
\oint x\ dp &=&\frac{2}{m\omega}\int_{-z}^{z}\frac{\sqrt{z^2-p^2}}{\bigl[1- 2\alpha p + q \alpha^2 p^2 \bigr]} dp \label{semi}\\
&=&\frac{2 z^2}{m\omega}\int_{-1}^{1}\frac{\sqrt{1-y^2}}{\bigl[1- 2\alpha zy + q \alpha^2 z^2 y^2\bigr]} dy \label{semi1}
\end{eqnarray}
where $\pm z$ are the classical turning points in the first integral. The second integral was obtained by scaling, $p=zy$.  

The integral (\ref{semi1}) may be evaluated exactly and is discussed in Appendix B; in particular the results for $q=0$ are identical to those obtained by solving the Schrodinger equation! Here however we would like to illustrate a procedure for isolating the leading correction to the bound state energies due to the deformation parameter $\alpha$. Such an approximate evaluation will be useful later when we extend the discussion to other potentials in subsequent sections.

%&=&\frac{2}{m\omega}\int_{-z}^{z}\Bigl(1+2\alpha p+(4-q)\alpha^2 p^2+4(2-q)p^3\alpha^3 + ...\Bigr)\sqrt{z^2-p^2}\label{expan} \\
%&=&\frac{\pi E^{sc}\bigl(2 + m(4-q)\alpha^2 E^{sc}\bigr)}{\omega}\equiv\bigl(n+1/2\bigr)h.\label{38a}

Notice that in the semi-classical expression (\ref{semi1}) the effects of the deformation reside solely in the weight factor $1/f$. As discussed earlier, for $q<1$, $f(p)$ has real zeros and hence for the integrals to be well-defined, a cut-off must be imposed on the momenta so that poles of $1/f$ are outside the integration limit in (\ref{semi}). This implies a restriction on the upper limit of integration $z$, and hence since $z=\sqrt{2mE}$, a maximum energy to the bound state spectrum; for $q=0, \ 2 \alpha z < 1$ and for $q=1, \ \alpha z < 1$. For $q>1$ there is no restriction. 

Now for $\alpha z \ll 1$ we may expand  
\begin{equation}
{1 \over f(p)} = 1 + 2 \alpha z y + (4-q) \alpha^2 z^2 y^2 + O(\alpha z)^3 \, . \label{expanS}
\end{equation}
 By symmetry the $O(\alpha)$ term does not contribute to (\ref{semi1}) and so 
 \begin{equation}
 { m\omega \over 2} \oint x\ dp = z^2 [ I_{01} + (4-q) (\alpha z)^2 I_{11} ] \label{phase}
 \end{equation}
 where we have defined the positive quantity
 \begin{eqnarray}
 I_{\tau \sigma} &\equiv & \int_{-1}^{1} y^{2\tau} (1-y^2)^{1 \over 2 \sigma} \  = \ B\Big({2\tau+1 \over 2}, {1 + 2 \sigma \over 2\sigma} \Big) \,,
 \end{eqnarray}
 $B(a,b)$ being the usual Beta function.
 
Thus inverting (\ref{phase}) gives,
\begin{equation}
z^2 = {m \omega h \over 2 I_{01}} (n + 1/2) \Big[ 1 - (4-q) (\alpha z)^2 {I_{11} \over I_{01}} \Big] + O(\alpha^3)   \ .
\end{equation}
This last equation can be solved by iterating around $\alpha=0$,
\begin{equation}
E^{sc}_{n} = E^{sc}_{n (0)} \Big[ 1 -  (4-q) (2m \alpha^2) {I_{11} \over I_{01}} E^{sc}_{n (0)} \Big] +O(\alpha)^3 \ 
\end{equation}
where the undeformed energies are given by
\begin{equation}
  E^{sc}_{n (0)} = {h  \omega \over 4  I_{01}} (n + 1/2)  = \hbar \omega (n + 1/2) \, . 
\end{equation}
Thus the energy shifts for $\alpha z \ll 1$ and $n$ large, 
\begin{eqnarray}
\Delta E^{sc}_{n} &=&  -  (4-q) (2m \alpha^2) {I_{11} \over I_{01}} (E^{sc}_{n (0)})^2  + O(\alpha^3 )  \, , \label{semihar} \\
&=& -  \frac{(4-q) m \alpha^2}{2} (E^{sc}_{n (0)})^2 \, ,
\end{eqnarray}
are negative for $q<4$, which includes the case $q=1$ corresponding to the exactly realised MCR (\ref{exact}). One can check that this expression agrees with the $O(\alpha^2)$ terms of the exact expressions in Sect.(4), bearing in mind that the semi-classical evaluation (at leading order in the $\hbar$ expansion) is expected to hold for large $n$ (and so, for example, terms independent of $n$ might not show up fully). The expression (\ref{semihar}) may also be obtained after an  exact evaluation of the  integral (\ref{semi1}), see Appendix B.

\subsection{Power Law Potentials}
We can easily extend the semi-classical analysis to symmetrical power law potentials of the form,
\begin{eqnarray}
V(X)&:=& {\lambda \over 2m} X^{2\sigma}\label{power}
\end{eqnarray}
where $0<\sigma < \infty$ and $\lambda$ is independent of $X$.  For instance, $\sigma=1$ and $\lambda = (m \omega)^2$ for the harmonic oscillator\footnote{The $\sigma \to \infty$ limit corresponds to the infinite well which will be discussed in detail in \cite{prep}.}. 

As before,
\begin{equation}
z^2 = 2m E^{sc} = P^2 + \lambda X^{2\sigma} \nonumber
\end{equation}
and so
\begin{equation}
X = \lambda^{-1/2\sigma} (z^2 -p^2)^{1/2\sigma} \label{power2}
\end{equation}
from which the canonical coordinate $x=X/f$ follows. The rest of the steps are identical to those discussed for the harmonic oscillator so we simply present the result. For $\alpha z \ll 1$ and large $n$,
\begin{eqnarray}
\Delta E^{sc}_{n} &=&  -  (4-q) (2m \alpha^2) {I_{1\sigma} \over I_{0\sigma}}  { 2\sigma \over \sigma +1} (E^{scp}_{n (0)})^2  + O(\alpha)^3 \label{powercorr} \\
& \Rightarrow & \frac{\Delta E^{sc}_{n}}{E^{scp}_{n (0)}}  \propto  -(4-q) E^{scp}_{n (0)} \, ,
\end{eqnarray}
with \cite{nieto}
\begin{equation}
E^{scp}_{n (0)} = {\lambda^{1 \over \sigma +1} \over 2m} \Big[{  (n + 1/2) h \omega \over 2  I_{0\sigma}} \Big]^{2\sigma \over \sigma +1} \ . \label{unpertP}
\end{equation}

Notice that for all power law potentials (\ref{power}), the leading $O(\alpha^2)$ correction to the energy at large $n$ is negative for $q <4$.  

The expression (\ref{powercorr}) holds for $\alpha z \ll 1$. One may also evaluate the semi-classical integrals for 
$\alpha z \gg 1$. Consider
\begin{equation}
\int_{-1}^{1} \frac{ (1-y^2)^{1 \over 2 \sigma} } { (1- \alpha zy)^2  + (q-1) (\alpha z y)^2 } dy \ .  
\end{equation}
For $y \neq 0$ and $\alpha z \gg1$ the integral is of order $1/(\alpha z)^2$. In order to isolate the contribution near $y=0$, write the numerator of the integrand $(*)$ as  $[(*)-1] + 1$ and split the integral into two pieces. The first integral is now ``safe" at $y=0$ and gives a $O(1/(\alpha z)^2)$ contribution. In the second integral scale $y \to y/(\sqrt{q-1} \alpha z)$ for $q>1$ and take $\alpha z \gg1$ to get $\sim 1/(\sqrt{q-1} \alpha z)$. Hence the semi-classical quantisation gives for $q>1$ and $\alpha z \gg 1$
\begin{equation}
E^{sc} \sim (\sqrt{q-1} \alpha n )^{2\sigma} \ . \label{vlarge}
\end{equation}
This result (\ref{vlarge}) may also be obtained by evaluating the integrals exactly and taking limits, see Appendix B. Notice that for the harmonic oscillator, $\sigma =1$, the large $n$ eigenvalues for $q>1$ approach the infinite well form, as remarked in Sect.(4.1)

\subsection{Comparison with string-motivated MCR}

As mentioned earlier, the MCR (\ref{string1}) has been used to solve exactly for the spectrum of some one dimensional problems. In order to uncover general trends, we can use a semi-classical analysis as in the last section  to study again power law potentials of the form (\ref{power}) but for the MCR (\ref{string1}). Since $\beta >0$, for large $n$ but $\beta z^2 \ll 1$ it is easy, by examining the weight factor, to deduce that the energy correction is now always positive,
 \begin{equation}
\Delta E^{sc}_{n} \propto + \beta (E^{scp}_{n(0)})^{2} \, , \label{powerE}
\end{equation}
while for  $\beta z^2 \gg 1$ the energy is similar to (\ref{vlarge}) but with $\alpha \sqrt{q-1}$ replaced by $\sqrt{\beta}$.  (Alternatively, as mentioned after Eq.(11), the string MCR results may also be obtained by taking $\alpha \to 0$ and $q \alpha^2 \to \beta$ in the results of the previous section.)

In other words, for the string-theory motivated MCR (\ref{1}) our semi-classical analysis suggests that for power-law potentials the deformed energy spectrum lies above the undeformed case. This is consistent with the exact solution for the harmonic oscillator and infinite wells in \cite{Pedram3}. By contrast, the  MCR's (\ref{10d}) with $q<4$ give negative energy corrections at intermediate energies. 

\subsection{Other MCR's}
There is no fundamental reason to limit oneself to quadratic momentum terms on the right-hand-side of relations such as (\ref{10d},\ref{string1}). One interesting proposal \cite{Pedram4} is 
\begin{eqnarray}
[X, P]&=& i\hbar \tilde{f}(p)  \label{ped} \, 
\end{eqnarray} 
with $ \tilde{f} = {1/( 1-\beta P^2)}$ which for small $\beta P^2$ agrees with (\ref{string1}) but through the pole at $P = 1/ \beta$ suggests a maximum momentum. A semi-classical analysis for the harmonic oscillator in Ref.\cite{Pedram4} showed the existence of a maximum bound state energy. 

We wish to show here, again by means of a semi-classical analysis, that a maximum bound state energy due to the MCR (\ref{ped}) is suggested for all power-law. Using similar notation as before,  we obtain
\begin{eqnarray}
(n + 1/2)h &=& z^{1+ 1/\sigma} \int_{-1}^{1} (1-y^2)^{1 \over 2\sigma} ( 1- \beta z^2 y^2) dy \nonumber \\
& \equiv & z^{1+ 1/\sigma} ( c_{1\sigma} - \beta c_{2\sigma} z^2)  \label{nonpol}
\end{eqnarray}
where $c_{1\sigma}, c_{2\sigma}$ are positive constants, and where we have suppressed some other positive constants. Note that unlike the analysis in previous sections, the second term on the right in Eq.(\ref{nonpol}) is complete rather than a leading term of a $\beta$ expansion\footnote{Of course being a leading semi-classical result, it is still not the same as the exact energy expression.}. Clearly then, if the energy ($E \sim z^2$) were to increase without bound the right-hand-side of (\ref{nonpol})  would become negative in contradiction to the left side which is always positive. Thus we conclude that there is a maximum energy in the bound state spectra of all power law potentials with the MCR (\ref{ped}). Since the maximum $z$ is clearly $O(\beta^{-1/2})$, the maximum energy is  $O(1/\beta)$ and the maximum number of bound states is  $n_{max} \sim O(1/ \beta^{(1 + 1/\sigma)/2})$. 

Other examples of MCR's  which involve a momentum cutoff are $\tilde{f} = 1 -\beta P^2$ \cite{mig} or $\tilde{f} = \sqrt{1 -\beta P^2}$ \cite{cutoff}. In a semi-classical analysis these result in a pole in the weight factor and thus a maximum energy to the bound state spectrum.

\subsection{Concave Potentials and Relativistic Dispersion}
Given the results for power-law potentials, it is intuitively reasonable to expect that similar conclusions hold for more general concave potentials in one dimension. Recall that in the semi-classical limit, it is the weight factor $1/f(p)$ in the integrals which is solely responsible for the deformation.

Consider a symmetric concave potential, $V(x)=V(-x), V(0)=0$ such that for $x>0, V(x_2) > V (x_1)$ if $x_2 > x_1$. For $x>0$ define the inverse function $U$ by $U(V(x)) = x$; for example for $V(x) =x^2, U(y) = \sqrt{y}$. For brevity, we choose the mass normalisation here such that $2m=1$. Then the integral in (\ref{semi}) is replaced by 
\begin{equation}
J(\alpha,q,z) \equiv \int^{z}_{-z}dp\frac{ U(z^2-p^2)}{\Bigl[1- 2\alpha p + q \alpha^2 p^2\Bigr]} \nonumber 
\end{equation}
For $\alpha z \ll 1$ we expand the denominator as before, giving 
\begin{equation}
J(\alpha,q,z) = J(0,q,z) + (4-q) \alpha^2 J_1(z) + O(\alpha^4) \, . \nonumber
\end{equation}
where the integrals $J(0,q,z), J_1(z) >0$.  Thus for large $n$ and up to $O(\alpha^2)$, the semi-classical quantisation condition can be written
\begin{equation}
E^{sc} = z^2 \sim { n h \over J(0,q,z)/z^2} \Big( 1- (4-q)\alpha^2  {J_1 (z) \over J(0,q,z) } \Big)  \, . \nonumber
\end{equation}
This equation may be solved (implicitly) by iteration about $\alpha=0$ giving
\begin{equation}
E_{n}^{sc} \sim E_{n(0)}^{scc}  \Big( 1 + (q-4)\alpha^2  {J_1 (z_0) \over J(0,q,z_0) } \Big)  \label{concave}
\end{equation}
where $E_{n(0)}^{scc}=z_{0}^{2}$ is the unperturbed  semi-classical energy obtained by solving $ n h= J(0,q,z_0)$. As before the energy correction is negative for $q<4$ and a similar analysis easily shows that the string-motivated MCR (\ref{string1}) still gives a positive correction for general concave potentials. 

Hence, a result of Sections (4.1,4.2) comparing the MCR's (\ref{10d},\ref{string1}) for power-law potentials holds more generally for symmetric concave potentials: For intermediate energies, (\ref{10d}) gives a negative energy correction for $q<4$ whereas the string theory motivated MCR gives a positive correction. 

Likewise, the expression (\ref{nonpol}) is easily generalised for symmetric concave potentials, again implying a maximum energy for the MCR (\ref{ped}). 

So far our discussion has been for the non-relativistic dispersion relation $E = P^2 +V(X)$. If the energies are so large that one is in the relativistic regime then in $c=1$ units the appropriate dispersion relation is $(E-V)^2 = p^2 + m^2$ giving for particles 
\begin{equation}
E = +\sqrt{p^2 + m^2} +V \, \label{rel} \, .
\end{equation}
We assume that the particles only interact through a potential $V$ which is weak, $|V| < 2m $, and slowly varying over the scale of the particle's Compton wavelength. Then pair production is not enabled and multi-particle effects can be ignored. The use of (\ref{rel}) in previous semi-classical integrals changes the magnitude of energy shifts due to the deformation but not their sign which is determined by the weight factor. Some explicit relativistic calculations with MCR's of the type (\ref{string1},\ref{adv})  are in \cite{rel-string,das4}.

\section{Higher dimensions}
In $D-$dimensional space, the momentum space representation of the algebra (\ref{commute},\ref{exact}) is
\begin{eqnarray}
P_i &=& p_i \\
X_{i}&=& i\hbar (1-\alpha p) \left[\frac{\partial}{\partial p_i}-\alpha {p_{i}p_{j} \over p} \frac{\partial}{\partial p_j} \right] \label{Ddim} \ .
\end{eqnarray}
Hermiticity of $X_i$ now requires the weight factor $(1-\alpha p)^{-(D+1)}$ in momentum integrals such as (\ref{16}). 

\subsection{Isotropic Oscillator in Semi-classical Limit}

Consider the $X^2 \psi(p)$ term in the momentum space Schrodinger equation,
\begin{equation}
X^2 \psi(p)= (i\hbar)^2 (1-\alpha p)^2 \left[\frac{\partial}{\partial p_i}-\alpha {p_{i}p_{j} \over p} \frac{\partial}{\partial p_j} \right]^2 \psi(p) \label{Dsq} \ , \nonumber
\end{equation}
repeated indices being summed. We wish to solve the equation in the WKB limit, $\hbar \to 0$ \cite{bender}. Write as usual $\psi(p) = \exp{(-i W(p) / \hbar)} $ with $W(p) = W_{0}(p) + \hbar W_{1} (p) + ...$ in a $\hbar$ expansion. Since $X^2$ comes with the 
$(\hbar)^2$ piece, thus derivatives acting on the wavefunction must pull down at least $1/\hbar^2$ factors for the net contribution to survive the $\hbar \to 0$ limit. This means that each derivative must act directly on $W(p)$:
\begin{equation}
{X^2 \psi(p) \over \psi(p)} \to (1-\alpha p)^2 \left[(\partial_i W)^2- {2\alpha \over p} (p_{i} \partial_i W) (p_{j} \partial_j W) + \alpha^2 (p_j \partial_j W)^2 \right] \label{Dsq2} \ . \nonumber
\end{equation}
We shall further restrict our attention to states without angular dependence, the $S$-states. For such wavefunctions,
\begin{equation}
p_i \partial_i W(p) = p \partial_{p} W(p) \nonumber
\end{equation}
and so
\begin{equation}
X^2 \psi(p) \to (1-\alpha p)^4 (\partial_p W)^2 \psi(p) \label{Dsq3} \ . \nonumber
\end{equation}
The time-independent Schrodinger equation for the harmonic oscillator is thus reduced to the relation
\begin{equation}
E^{sc} = {p^2 \over 2m} + {m \omega^2 \over 2} (1-\alpha p)^4 (\partial_p W)^2 \ . \label{scd}
\end{equation}
From here one may, although we will not, proceed with a standard WKB type analysis to obtain the quantisation condition
\begin{eqnarray}
\oint x\ dp=\bigl(n+ \gamma) h; \hspace{0.35cm} n=0,1,2,3....\label{BS2}
\end{eqnarray} 
where $\gamma$ is a constant which depends on boundary conditions (and thus also on space dimension $D$).

We will instead motivate the quantisation condition through the correspondence principle. Writing $W = \int^{p} \chi(p') dp'$, (\ref{scd}) becomes
\begin{equation}
E^{sc} = {p^2 \over 2m} + {m \omega^2 \over 2} (1-\alpha p)^4 (\chi)^2 \ . \nonumber
\end{equation}
One can identify $\chi$ by considering the limit $\alpha \to 0$ whereby it is clear that $\chi$ is $x$, the canonical classical space coordinate. As is well known \cite{messiah}  for $\alpha =0$,  the phase space area $\oint x\ dp = \oint p\ dx$ is an adiabatic invariant for bounded orbits in the classical theory, and is thus naturally identified with stationary states in the quantum theory through the condition
\begin{eqnarray}
{ 1 \over 2} \oint x\ dp= {n h \over 2} ; \hspace{0.35cm} n= 1,2,3....\label{BS3}
\end{eqnarray}
For $\alpha \neq 0$, $\oint x\ dp$ will be an adiabatic invariant of the deformed classical dynamics and (\ref{BS3}) is then the semi-classical quantisation condition in the deformed quantum theory. (Indeed, as the canonical relation $[x,p]=i\hbar$ goes over to the Poisson bracket, the deformed commutator $[X,P]$ goes over into the deformed Poisson bracket). 

Recall the physical meaning of (\ref{BS3}) in the quantum theory: It counts the number of half-deBroglie wavelengths that fit into the potential well at a particular energy. The difference between (\ref{BS2}) and (\ref{BS3}) is that the former takes into account quantum ``leakage" at the ends. This difference is immaterial at large $n$ which we focus on here. 

Thus for the $S$-states of the $D$-dimensional isotropic oscillator, the semi-classical bound state spectrum is determined by
\begin{eqnarray}
n h &=&\frac{2}{m\omega}\int_{o}^{z}\frac{\sqrt{z^2-p^2}}{(1- \alpha p)^2} dp \label{semiD}
\end{eqnarray}
which should be compared to the one-dimensional oscillator relation (\ref{semi}): The weight measure is precisely that for one-dimension, and not the $D$-dimensional weight mentioned after Eq.(\ref{Ddim}). The  intrinsic momentum cut-off $p < 1/\alpha$ is also implied. 

However note that the coordinate $p$ here is the radial momentum coordinate and so the lower limit of the integral is zero as opposed to the one-dimensional case where $p$ was on the open line. Recalling the discussion following Eq.(\ref{expanS}), this means that for this higher dimensional oscillator the leading correction will be  $O(-\alpha)$ and will be negative for $\alpha >0$. 

\subsection{Spectrum of Isotropic Power-Law Potentials}
A generalisation to $S$-states of $D$-dimensional radial potentials of the form 
\begin{eqnarray}
V(X)&=& {\lambda \over 2m} X^{2\sigma} \nonumber
\end{eqnarray}
is straightforward in the semi-classical limit as 
\begin{eqnarray}
X^{2\sigma} \psi(p) &=& (X^2)^{\sigma} \psi(p) \to [(1-\alpha p)^4 (\partial_p W)^2]^{\sigma} \psi(p)  \ . \nonumber
\end{eqnarray}
The rest of the discussion is as in the previous subsection, with expressions similar to the one dimensional case, for example (\ref{power2}), but with $q=1$ and the lower limit of the integral being zero. The leading correction is again $O(-\alpha) < 0 $ for $\alpha >0$. Although our discussion so far has been for the exact MCR (\ref{exact}), the conclusion 
holds for the class of MCR's (\ref{approxquad}) since they share the same leading $O(-\alpha)$ term. 

\subsection{String-theory Motivated MCR}
To check the accuracy of the semi-classical analysis above, we apply it to the $D$-dimensional version of the string-motivated algebra Eq.(\ref{string1}), 
\cite{kempf,laynam},   
\begin{eqnarray}
[X_{i},P_{j}] &=& i\hbar \Bigl( \delta_{ij} + \beta \delta_{ij} P^{2} +\beta^{'}P_{i}P_{j} \Bigr)  \label{stringD}   \ ,  \\
\left[X_{i}, X_{j} \right] &=& i\hbar \frac{ (2\beta -\beta^{'}) + (2\beta + \beta^{'}) \beta P^2}{1+ \beta P^2} (P_i X_j - P_j X_i)  \ , \\
\left[P_{i},P_{j} \right] &=& 0 , 
\end{eqnarray}
which is realised in momentum space by
\begin{eqnarray}
P_i &=& p_i \nonumber \\
X_{i}&=& i\hbar (1 +\beta p^2) \frac{\partial}{\partial p_i} +\beta^{'} p_{i} p_{j} \frac{\partial}{\partial p_j}  \nonumber \ .
\end{eqnarray}
We note in passing that For $\beta=0$ (\ref{stringD}) is the non-relativistic version of Snyder's algebra \cite{snyder}. 

In the WKB limit, one finds that for $S$-states, 
\begin{equation}
X^2 \psi(p) \to [(1 + (\beta + \beta^{'}) p^2 ]^2 (\partial_p W)^2  \psi(p) \nonumber \ 
\end{equation}
and hence for the $D$-dimensional oscillator the expression is simply (\ref{semiD}) but with the weight factor replaced by $(1 + (\beta + \beta^{'}) p^2)^{-1}$. Explicitly, we find
\begin{eqnarray}
\Delta E^{sc}_{n} &=&  \frac{(\beta + \beta^{'}) m }{2} (E^{sc}_{n (0)})^2 + O(\beta + \beta^{'})^2   \label{powerstring}
\end{eqnarray}
which agrees at large $n$ and leading order with the exact expressions in Ref.\cite{kempf,laynam}. The correction (\ref{powerstring}) is 
is positive since $\beta + \beta^{'} > 0$ is typically assumed for the  MCR (\ref{stringD}) in order to reproduce a minimum position uncertainty \cite{kempf}.

The generalisation to isotropic power law potentials again essentially involves replacing the weight factor in previous one-dimensional results; the net result is  a positive correction of $O(\beta + \beta^{'})$.

\section{Conclusion}

We obtained a two-parameter class of exactly realised MCR's which implied an intrinsic maximum momentum, a subset of which also implied a minimum position uncertainty (\ref{F2},\ref{G1}). Among the exactly realised MCR's a one-parameter MCR (\ref{exact}) had only linear and quadratic momenta on the right. 

A two-parameter family of approximate MCR's was then constructed (\ref{approxquad}) from the exact solutions using a $(\alpha p)^2 \ll 1$ expansion; they satisfy the Jacobi identity approximately when $|r| (\alpha p)^2 \ll 1$ and differ from each other at terms of order $(\alpha p)^2$. The proposal of ADV \cite{das2} is an example from this class. 

For dynamics restricted to one dimension, the exact (\ref{exact}) and approximate MCR's were described by (\ref{10d}), with one deformation parameter $\alpha$ and a dimensionless parameter $q$. The exactly realised MCR has $q=1$ while the most studied version in the literature is the ADV proposal $q=4$. 
 
We studied the impact of maximum momentum by determining the bound state spectrum of the deformed harmonic oscillator exactly for $q=0$ and $q=1$; the spectra terminate at finite energy beyond which is the continuum, in contrast to the usual undeformed oscillator which has a purely discrete spectrum with no maximum energy. 

For both $q=0$ and $q=1$ we computed the deformed position and momentum uncertainties and their product $\Delta X \Delta P$ in an eigenstate $n$: The values were all lower than the $\alpha=0$ case, increasing with $n$ up to some maximum before declining to vanish at $n_{max}$. Thus the upper-most bound states display classical characteristics. 

%In Sect.(4.2.2) we conjectured that more generally, for MCR's of the form (\ref{class}) which realise an intrinsic maximum momentum, there exists superposition of states for which $\langle P^2 \rangle \sim p_{max}^2$ and $(\Delta P)^2 /  \langle P^2 \rangle \approx 0$.   

In the second part of the paper we used a semi-classical analysis, which we motivated in detail in Sect.(6), to study bound state spectra of general concave potentials in one dimension.   For $\alpha p \ll 1$ and large $n$, the energy shifts due to the deformation are negative for $q<4$ (\ref{concave}). Under similar conditions the string-motivated MCR (\ref{1}) gives a positive correction. Explicit expressions for the energy shifts were obtained for power-law potentials (\ref{powercorr}).
It is interesting to note that the ADV proposal $q=4$ gives a vanishing leading order correction at large $n$. 

Empirically, the conditions $ \alpha p \ll 1$ and $\beta p^2 \ll 1$ are natural since quantum mechanics has been very well tested, placing strong limitations on the suggested deformations in (\ref{string1}) and ({\ref{10d}) \cite{das1,optics}.  At the same time, those conditions allow one to ignore higher order (cubic) corrections to those MCR's and also enabled the extraction of leading order effects from the semi-classical integrals.  

The large $n$ condition was required for the validity of the semi-classical analysis, but since $\frac{\Delta E_n}{E_n} \propto E_n$ for large $n$, this regime is also physically relevant. In Sect.(5.5) we explained why relativistic corrections do not change the sign of the leading order energy shifts if one remains in the one particle sector. 

Thus one way that string-motivated  MCR's such as (\ref{string1},\ref{stringD}) can be distinguished from the $q<4$ sub-class of MCR's such as ( \ref{exact}, \ref{approxquad}) is through the sign of energy shifts in bound state spectra of dynamics restricted to one dimension. The Penning trap is one possible arena for such dynamics \cite{laynam}.

Note that the $q<4$ sub-class still includes $q>1$ for which the one-dimensional MCR's were low-momentum expansions of exact MCR's which implied a maximum momentum (through poles) and also a minimum position uncertainty. 

We also studied the $D$-dimensional $S$-states of isotropic power-law potentials semi-classically. For the string-motivated MCR (\ref{stringD}) we extended a result of \cite{laynam}, showing that $\Delta E \propto \beta >0$. For the MCR's (\ref{exact},\ref{approxquad}) the shift $\Delta E = O(-\alpha)$ is of opposite sign for the usually assumed $\alpha >0$.  Since $\beta p^2 \sim (\alpha p)^2 \ll 1 $, then regardless of the sign of $\alpha$,  MCR's such as (\ref{exact},\ref{approxquad}) may show experimental signatures earlier than the string suggested modifications (\ref{stringD}), if the MCR's are empirically realised.     

In this paper we studied various aspects of the MCR's (\ref{exact},\ref{approxquad}) in the momentum representation. Other aspects and applications of the MCR's obtained in this paper, such as a coordinate space representation and scattering problems, will be discussed elsewhere \cite{prep}.

\section{Acknowledgment}
R.P thanks Saurya Das and Pouria Pedram for stimulating discussions.

\section*{Appendix A}

Here we implement the Jacobi identity constraint exactly on the ansatz (\ref{1}, \ref{commute}). After some algebra one obtains
\begin{eqnarray}
0& \equiv &\hbar^{2}\left\{(\alpha_{1}-\alpha_{2})P^{-1}+(\alpha_{1}^{2}+2\beta_{1}-\beta_{2})+(3\alpha_{1}+\alpha_{2})\beta_{1}P\right.\nonumber\\
&\phantom{a}&\hspace{0.75cm}\left.+\beta_{1}(2\beta_{1}+\beta_{2})P^{2}\right\}\Delta_{jki}, \label{jacob}
\end{eqnarray}
where $\Delta_{jki}:= P_i \delta_{jk} - P_j \delta_{ik}$.

For $D>1$ dimensions the coefficient of each power of $P$ inside the curly brackets must vanish, so  the parameters $(\alpha, \beta)$ satisfy four simultaneous equations, 
\begin{eqnarray}
\alpha_{1}-\alpha_{2}&=&0 \label{a1} \ ,\\
\alpha_{1}^{2}+2\beta_{1}-\beta_{2}&=&0 \label{a2} \ ,\\
\beta_{1}(3\alpha_{1}+\alpha_{2})&=&0 \label{a3}  \ ,\\
\beta_{1}(2\beta_{1}+\beta_{2})&=&0 \ .   \label{a4}
\end{eqnarray}

The only nontrivial solution is given by
\begin{eqnarray}
\alpha_{1}&=&\alpha_{2}:=-\alpha\\
\beta_{1}&=&0\\
\beta_{2}&=&\alpha^{2}_{1}=\alpha^{2}, \label{9}
\end{eqnarray}
from which we infer that the  deformed commutator (\ref{1}) must be
\begin{eqnarray}
[X_{i},P_{j}]&=& i\hbar (1-\alpha P) \left[\delta_{ij}-\alpha {P_{i}P_{j} \over P} \right] \label{exactA} \ ,
\end{eqnarray}
which is precisely the expression (\ref{exact}) obtained in Sect.(2) from the class (\ref{class}). 

In Ref.\cite{das2} the Jacobi identity constraint on (\ref{1}) was first solved to leading order in $\alpha$, obtaining only Eqs.(\ref{a1},\ref{a2}). The choice $\beta_{1}\approx\alpha^2$ was then made leading to $\beta_{2}\approx3\alpha^2$ and the commutator
\begin{eqnarray}
[X_{i},P_{j}]_{\text{ADV}}& \approx & i\hbar\left[\delta_{ij}-\alpha\left(\delta_{ij} P+ \frac{P_{i}P_{j}}{P}\right)+\alpha^{2}(\delta_{ij}P^2+ 3P_{i}P_{j})\right]\label{adv}
\end{eqnarray}
In other words, in Ref.\cite{das2} the additional constraints (\ref{a3},\ref{a4}) were not implemented.

As discussed in Sect.(2), the ADV commutator belongs to the class of approximate commutators (\ref{approxquad}) which differ from each other at order $(\alpha p)^2$ and whose regime of validity is $ |r| (\alpha p)^2 \ll 1$.  For the one-dimensional projections (\ref{10d}), the ADV commutator corresponds to $q=4$.

\section*{Appendix B: Semi-classical Integrals}
The integral in (\ref{semi1}) can be evaluated exactly by passing to the complex plane. Denote $E_{n(0)} = (n + 1/2) \hbar \omega$ and $z^2 = 2mE^{sc}_{n}$.
Then
\begin{equation}
1+ \alpha^2 q m E_{n(0)} \equiv H(q, \alpha, z) \, . \label{master}
\end{equation}

There are three cases to consider. For $q < 1$ and $\alpha z < (1-\sqrt{1-q})/q$, we have  $H(q,\alpha,z)=H_1$ where
\begin{eqnarray}
H_{1} &\equiv & \frac{1}{2\sqrt{1-q}} (A-B) \ , \\
A \equiv [ (1 + \sqrt{1-q})^2 - (q\alpha z)^2 ] ^{1/2} \ , \nonumber \\
B \equiv [ (1 - \sqrt{1-q})^2 - (q\alpha z)^2 ] ^{1/2} \, . \nonumber
\end{eqnarray}

For $q>1$ and $\alpha z < 1/q$, $H= H_2$ with
\begin{eqnarray}
H_2 &\equiv & C_1 C_2 \\
C_1 & \equiv & \frac{1}{\sqrt{q-1}} [ (2-q- q^2 \alpha^2 z^2)^2 + 4(q-1) ] ^{1/4} \ , \nonumber \\
C_2 & \equiv & \frac{1}{\sqrt{2}} [ 1 - N_2/D_2 ]^{1/2} \ , \nonumber \\
N_2 &\equiv & 2-q- q^2 \alpha^2 z^2 \nonumber \\
D_2 &=& q [( 1 + 2(q-2)(\alpha z)^2 + q^2 (\alpha z)^4 ]^{1/2}\, . \nonumber 
\end{eqnarray}

For $q>1$ and $\alpha z > 1/q$, $H= H_3$ with
\begin{eqnarray}
H_3 & \equiv & C_1 C_3 \ , \\
C_3 & \equiv & \frac{1}{\sqrt{2}} [ 1 - N_3/D_2 ]^{1/2} \ , \nonumber \\
N_3 &\equiv & q- q^2 \alpha^2 z^2 \ , \nonumber
\end{eqnarray}
where $C_1$ and $D_2$ are the same as for the $H_2$ case. 

In the limit $q \to 1$, 
\begin{equation}
H_1 =H_2 =H_3 = \frac{1}{\sqrt{1-\alpha^2 z^2}} \ . 
\end{equation}
Notice the original constraint $\alpha z <1$, that is $E^{sc}_{n} \le E_{max}=1/(2 m \alpha^2)$, is contained in the expression. Solving (\ref{master}) for this case gives
\begin{equation}
E_{n}^{sc} = E_{n(0)} \left[ \frac{1+ {\alpha^2 m \over 2} E_{n(0)}}{ (1+ \alpha^2 m E_{n(0)})^2 } \right]
\end{equation} 
which shows that $E_{n}^{sc} \to E_{max} =1/(2m\alpha^2)$ as $n \to \infty$, the same limit attained by the exact expression in Sect.(4).

For the case $q \to 0$, 
\begin{equation}
H_1 \to 1 + {q \over 4} [ 1- (1-4\alpha^2 z^2)^{1/2} ]
\end{equation} 
Notice the original $2\alpha z <1$ constraint encoded inside $H_1$; the energy is bounded above: $E^{sc}_{n} \le 1/(8m \alpha^2)$. Solving (\ref{master}) in this case gives precisely the exact energy spectrum (\ref{32}).

For $\alpha z \ll 1 $ the energy shifts from all three cases above is summarised by (\ref{semihar}).

\newpage

\section*{Figure Captions}

\begin{itemize}
\item  Figure 1: The discrete spectrum for the $q=0$ oscillator. For this and the other figures $\alpha=0.1$ and $2m=\hbar \omega=1$. The straight line corresponds to $\alpha=0$.

\item  Figure 2: The probability density for $q=0, n=10$. The horizontal axis is $\xi =2(1-2\alpha P)/\delta^2$ and so $\xi=0$ corresponds to $p=p_{max}=1/(2 \alpha)$ while $\xi =\infty$ corresponds to $p= -\infty$.  

\item  Figure 3: The high-momentum $p$ (low $\xi$) end of the probability density for $q=0, n=49$. The horizontal axis in this and the next figure is labeled as in Fig.(2). However note the different scales in Figs.(2-4).

\item  Figure 4: The low-momentum $p$ (high $\xi$) end of the probability density for $q=0, n=49$.  

\item  Figure 5: The position uncertainty for the $q=1$ oscillator. 

\item  Figure 6 : The momentum  uncertainty for the $q=1$ oscillator. 

\item  Figure 7 : The uncertainty product for the $q=1$ oscillator. 

\end{itemize}

\section*{Figures}

\begin{figure}[ht]
  \begin{center}
   \epsfig{file=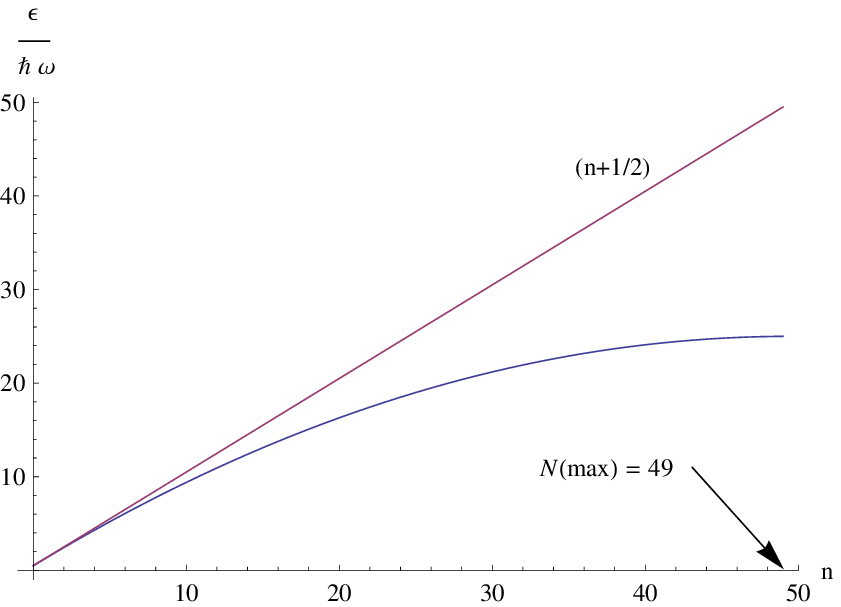, width=12cm}
    \caption{}
    %\label{e1}
  \end{center}
\end{figure}

\begin{figure}
  \begin{center}
   \epsfig{file=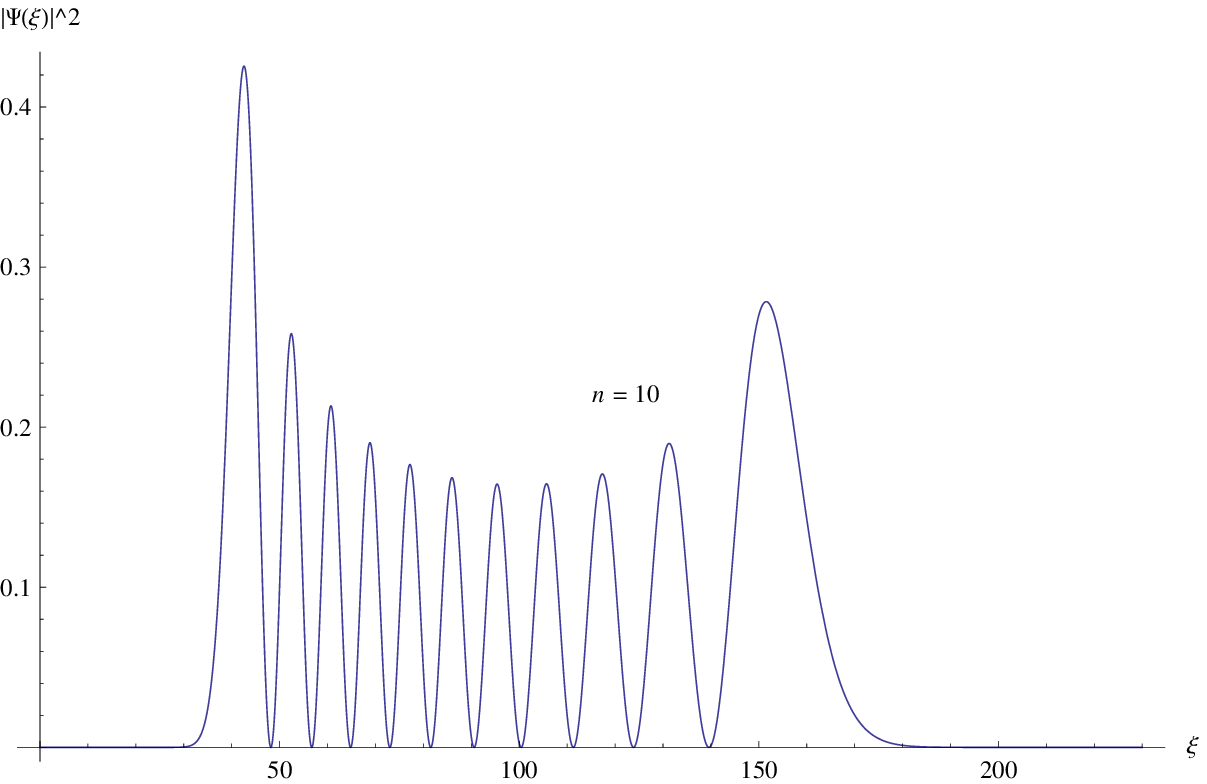, width=12cm}
    \caption{}
    %\label{e2}
  \end{center}
\end{figure}

\begin{figure}
  \begin{center}
   \epsfig{file=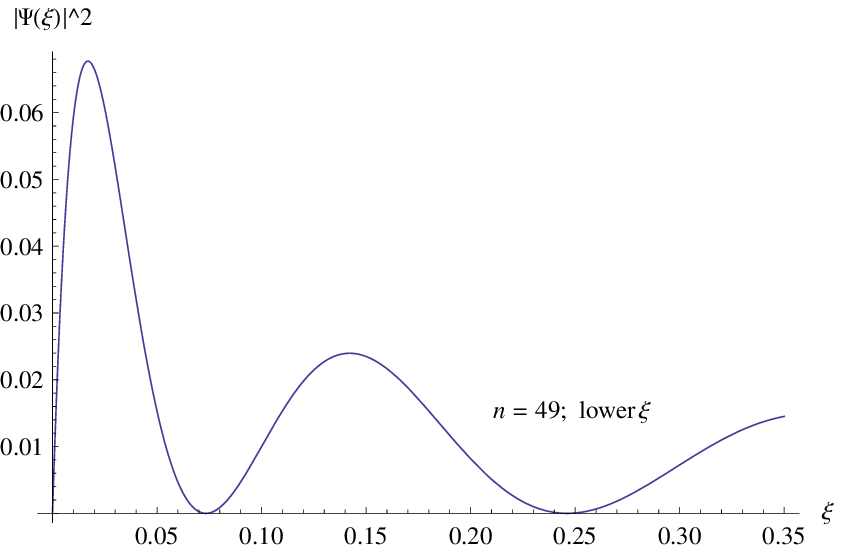, width=12cm}
    \caption{}
    %\label{e2}
  \end{center}
\end{figure}

\begin{figure}
  \begin{center}
   \epsfig{file=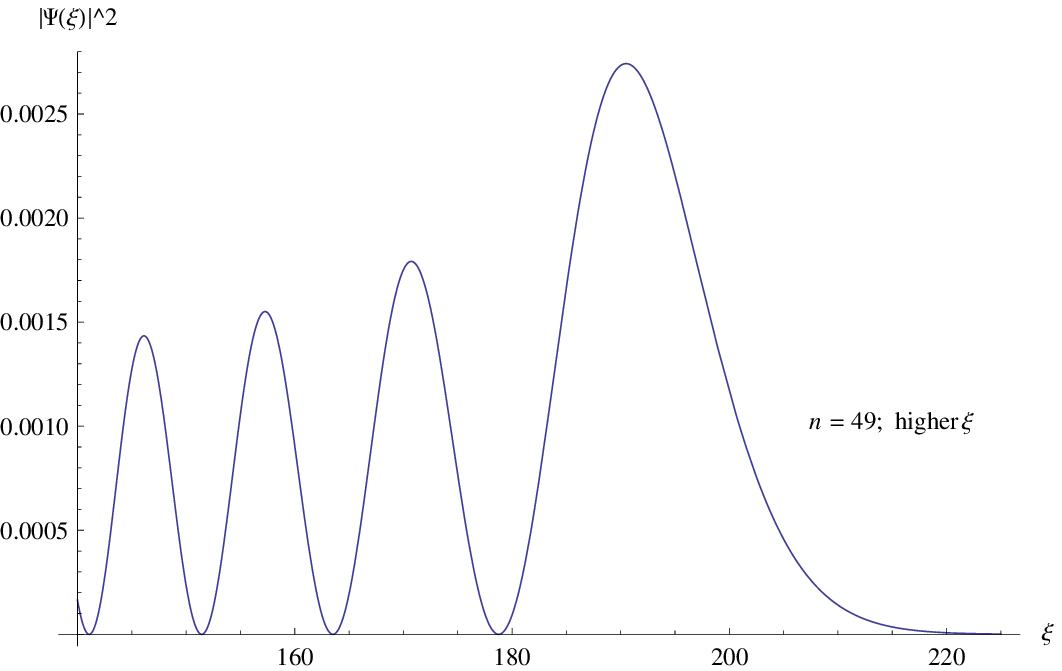, width=12cm}
    \caption{}
    %\label{e3}
  \end{center}
\end{figure}

\begin{figure}
  \begin{center}
   \epsfig{file=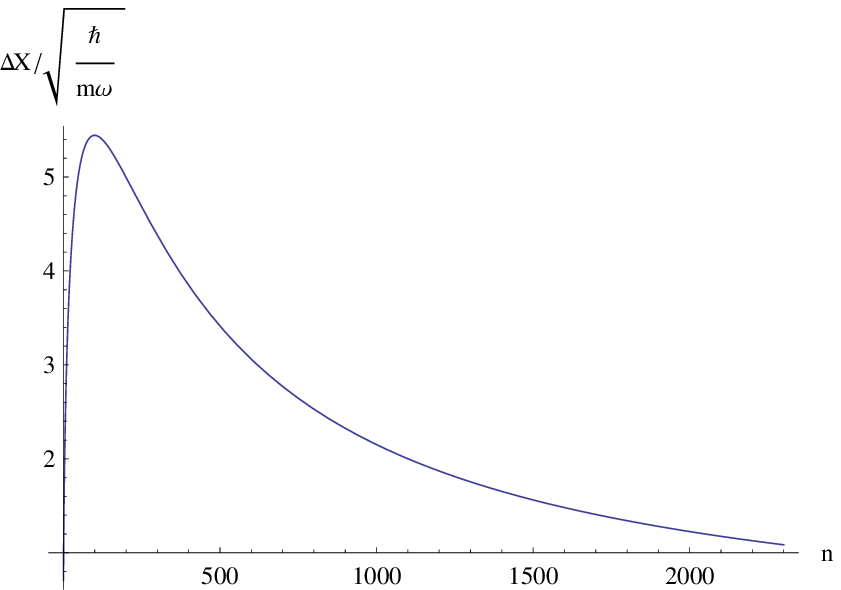, width=12cm}
    \caption{}
    %\label{e2}
  \end{center}
\end{figure}

\begin{figure}
  \begin{center}
   \epsfig{file=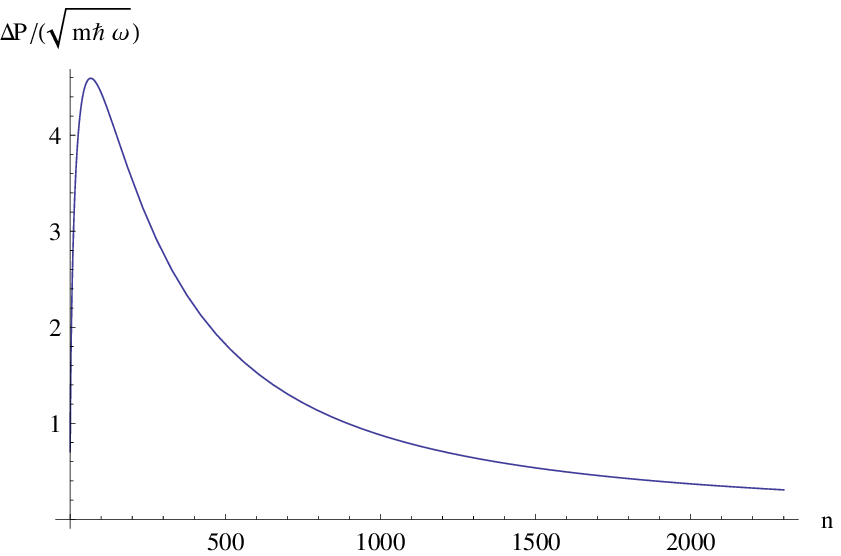, width=12cm}
    \caption{}
    %\label{e2}
  \end{center}
\end{figure}

\begin{figure}
  \begin{center}
   \epsfig{file=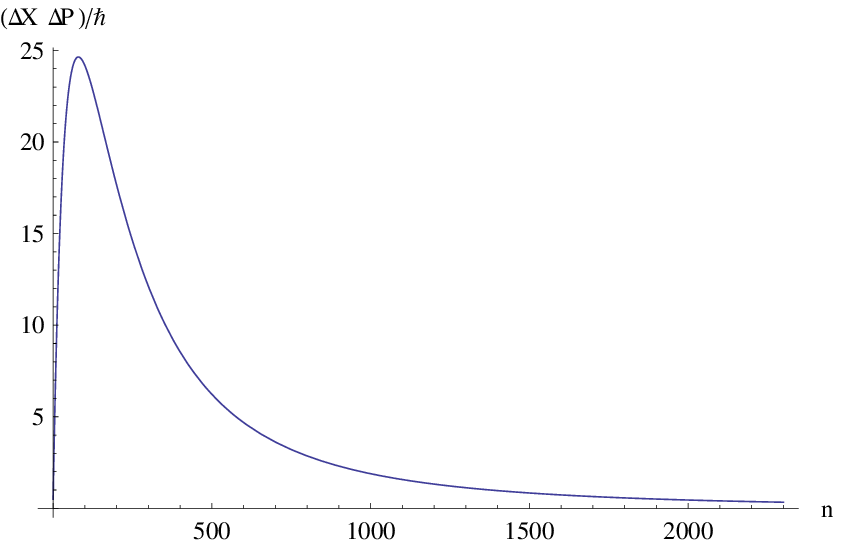, width=12cm}
    \caption{}
    %\label{e3}
  \end{center}
\end{figure}


\begin{thebibliography}{99}

\bibitem{min} L. J. Garay, Int. J. Mod. Phys. A {\bf 10} (1995) 145; R.A. Adler, Am. J. Phys. {\bf 78} (2010) 925. 

\bibitem{sabine} S. Hossenfelder, arXiv:1203.6191v1 [gr-qc].

\bibitem{string} D. Amati, M. Ciafaloni and G. Veneziano, Phys. Lett. B {\bf 216} (1989) 41; M. Maggiore, Phys. Lett. B {\bf 304} (1993) 65.

\bibitem{kempf} A. Kempf, G. Mangano and R.B. Mann, Phys. Rev. D {\bf 52} (1995) 1108; A. Kempf, J. Phys. A {\bf 30} (1997) 2093.

\bibitem{laynam} L.N. Chang, D. Minic, N. Okamura and T. Takeuchi, Phys. Rev. D {\bf 65} (2002) 125028; C. Quesne and V.M. Tkachuk, Phys. Rev. A {\bf 81} (2010) 012106; Z. Lewis and T. Takeuchi, Phys. Rev. D {\bf 84} (2011) 105029; L.N. Chang, Z. Lewis, D. Minic and T. Takeuchi, Advances in High Energy Physics (2011) 493514. 

\bibitem{das1} F. Brau, J. Phys. A {\bf 32} (1999) 7691; Akhoury R and Yao Y-P, Phys. Lett. B {\bf 572} (2003) 37; S. Das and E.C. Vagenas, Phys. Rev. Lett. {\bf 101} (2008) 221301; K. Nozari and P. Pedram, Euro. Phys. Lett. {\bf 92} (2010) 50013; ; D. Bouaziz and N. Ferkous,  Phys. Rev. A {\bf 82} (2010) 022105. 
%quadratic phenom


\bibitem{rel-string} U. Harbach, S. Hossenfelder, M. Bleicher and H. Stocker, Phys. Lett. B 584 (2004) 109;. C. Quesne and V.M. Tkachuk, J. Phys. A {\bf 38} (2005) 1747; K. Nouicer, J. Phys. A {\bf 39} (2006) 5125; Y. Chargui, A. Trabelsi, L. Chetouani, Phys. Lett. A {\bf 374} (2010) 531; M.Kober, Phys. Rev. D 82 (2010) 085017; P. Pedram, Phys. Lett. B {\bf 710} (2012) 478.
%rel string and adv gup

\bibitem{Pedram3} P. Pedram, Phys. Rev. D {\bf 85} (2012) 024016; P. Pedram, arXiv:1201.2802.
%nonpert-quad/sclass/exact. IJTP last bouncer

\bibitem{optics} I. Pikovski, M.R. Vanner, M. Aspelmeyer, M. Kim, C. Brukner, arXiv:1111.1979v1

\bibitem{brane} R. Maartens and K. Koyama, Living Rev. Relativity 13 (2010) 5.


\bibitem{dsr} J. Magueijo and L. Smolin, Phys. Rev. D {\bf 67} (2003) 044017; 
J.L. Cortes and J. Gamboa, Phys. Rev. D {\bf 71} (2005) 065015; G. Amelino-Camelia, Phys. Lett. B {\bf 510} (2001) 255. 


\bibitem{das2} A.F. Ali, S. Das and E.C. Vagenas, Phys. Lett. B {\bf 678} (2009) 497; A.F. Ali, S. Das and E.C. Vagenas, Phys. Rev. D {\bf 84} (2011) 044013.
%details, jocobi etc

\bibitem{das4} S. Das, E.C. Vagenas and A.F. Ali, Phys.Lett. B {\bf 690} (2010) 407; P. Pedram, Phys. Lett. B {\bf 702} (2011) 295.
%relativistic

\bibitem{Pedram1} P. Pedram, Euro. Phys. Lett {\bf 89} (2010) 50008.
%pert-das

\bibitem{Pedram2} P. Pedram, K. Nozari and S. H. Taheri, JHEP {\bf 03} (2011) 093.
%bouncer-das

\bibitem{Pedram4} P. Pedram, arXiv:1110.2999.
%higher order GUP. max mom

\bibitem{mig} S. Mignemi, Phys. Rev. D {\bf 84} (2011) 025021.

\bibitem{jiz} P. Jizba, H. Kleinert and F. Scardigli,  Phys. Rev. D {\bf 81} (2010) 084030. 

\bibitem{cutoff} M. Maggiore, Phys. Lett. B {\bf 319} (1993) 83; M.V. Battisti, Phys. Rev. D {\bf 79} (2009) 083506.




\bibitem{messiah} A. Messiah, Quantum Mechanics (Dover, 1999)

\bibitem{prep} C.L. Ching, et. al, in preparation. 

\bibitem{bender} C.M. Bender and S.A. Orszag, Advanced Mathematical Methods for Scientists and Engineers (Springer, 1999).



\bibitem{factor} F. Cooper F, A. Khare  and U. Sukhatme, Phys. Rep. {\bf 251} (1995) 267; 
C Quesne and V.M. Tkachuk, J. Phys. A {\bf 36} (2003) 10373. 



\bibitem{Fityo} T.V. Fityo, I.O. Vakarchuk, V.M. Tkachuk, J. Phys. A: Math. Theor. {\bf 41} (2008) 045305.






\bibitem{nieto} M.M. Nieto and L.M. Simmons, Jr., Am. J. Phys. {\bf 47} (1979) 634. 

\bibitem{snyder} H. S. Snyder, Phys. Rev. {\bf 71} (1947) 38.

\end{thebibliography}
\end{document}